\documentclass[11pt,letterpaper]{article}
\usepackage{soul}
\usepackage{jheppub}
\usepackage[mathletters]{ucs}
\usepackage[utf8]{inputenc}
\usepackage[T1]{fontenc}
\usepackage{amsfonts}
\usepackage{amsmath}
\usepackage{graphicx,subfigure}
\usepackage{tensor} 
\usepackage{mathtools} %
\allowdisplaybreaks[4]
\usepackage{amssymb}
\usepackage{amsthm,amsmath,amssymb}
\usepackage{mathrsfs}
\usepackage{euscript}
\usepackage{color}
\usepackage{braket}
\usepackage{orcidlink}

\title{ \boldmath Quasinormal modes of a charged spherically symmetric black hole in bumblebee gravity}

\author[a,b]{Bo-Rui Li,}
\emailAdd{320220902891@lzu.edu.cn}

\author[a,b]{Jia-Zhou Liu,}
\emailAdd{liujzh2025@lzu.edu.cn}

\author[a,b]{Wen-Di Guo\footnote{Corresponding author},}
\emailAdd{guowd@lzu.edu.cn}

\author[a,b]{Yu-Xiao Liu\footnote{Corresponding author}}
\emailAdd{liuyx@lzu.edu.cn}

\affiliation{$^{a}$Key Laboratory of Quantum Theory and Applications of MoE, Lanzhou Center for Theoretical Physics,
Key Laboratory of Theoretical Physics of Gansu Province, Gansu Provincial Research Center for Basic Disciplines of Quantum Physics, Lanzhou University, Lanzhou 730000, China\\
$^{b}$Institute of Theoretical Physics $\&$ Research Center of Gravitation,
School of Physical Science and Technology, Lanzhou University, Lanzhou 730000, China} 

\abstract{Recently,~exact charged spherically symmetric black hole solutions within the framework of bumblebee gravity have been obtained, where the Lorentz
symmetry is spontaneously broken due to the nonvanishing vacuum expectation
value of the bumblebee field. In this work, we investigate the quasinormal modes
of this black hole. We compute the quasinormal frequencies corresponding to the scalar perturbation and the gravito-electromagnetic coupled perturbation using both the
continued fraction method and the asymptotic iteration method. A detailed comparison of the results obtained from the two approaches is presented to
evaluate their accuracy and efficiency in this Lorentz-violating background.}

\begin{document}
\maketitle

\flushbottom


\section{Introduction}
General Relativity and Quantum Mechanics represent two fundamental pillars of modern physics. While General Relativity describes gravity and the large-scale structure of the universe, Quantum Mechanics governs the behavior of particles at microscopic scales. Despite their individual successes, these theories remain fundamentally incompatible, leading to ongoing efforts to develop a unified theory of quantum gravity. Current researches explore approaches, including string theory~\cite{Ooguri:1996ik, Matsunaga:2019fnc, Witten:1998qj, Witten:1998cd, Witten:1996hc, Witten:1995ex, Seiberg:1999vs} and loop quantum gravity \cite{Liu:2021djf, Bojowald:2022zog, Ziaeepour:2021ubo, Bojowald:2001ep, Fatibene:2024rwr, Rovelli:1997yv, Wang:2024jtp, Long:2021izw, Bianchi:2023avf, Duston:2011gk},~have been developed to reconcile these frameworks. 

Some current approaches to quantum gravity offer compelling theoretical frameworks.~However,~due to the limitations of present-day experimental precision,~these theories remain largely untestable.~From a theoretical perspective,~it is therefore essential to continue exploring and clarifying their fundamental properties,~while anticipating future advancements in experimental techniques that may eventually allow for their empirical verification.~In these theoretical models, the violation of Lorentz symmetry has become a major focus of research. Some disturbed strings are likely to break Lorentz symmetry in unstable vacuum environments~\cite{Kostelecky:1988zi, AraujoFilho:2024ykw, Kostelecky:2003fs}.~In 1989, Kostelecky and Samuel introduced the prototype of the bumblebee model~\cite{Kostelecky:1988zi}, a string-inspired framework featuring tensor-induced spontaneous Lorentz symmetry breaking.~In the framework of bumblebee gravity,~spontaneous Lorentz violation arises from a potential \(V(B^\mu B_\mu\pm b^2)\) that acts on a vector field \(B_\mu\)~\cite{Kostelecky:2003fs,Kostelecky:2010ze}. Within this model, numerous black hole solutions have been obtained ~\cite{Casana:2017jkc, Liu:2025oho, Maluf:2020kgf, Xu:2022frb, Ding:2019mal, Liu:2024axg, Ding:2023niy}, and the properties of these black holes have been extensively studied in various works~\cite{Kostelecky:2002hh, Bluhm:2007bd, Altschul:2009ae, Kozameh:2006hk, Ruffini:2002kz, Mattingly:2005re, AraujoFilho:2024iox, AraujoFilho:2024ctw, AraujoFilho:2025rvn, Shi:2025plr, Friedrich:2002xz, Kostelecky:2008ts, Guo:2023nkd, Zhao:2024uzq, Duan:2023gng, Liu:2025fxj, Kuang:2022xjp,Oliveira:2018oha,Gomes:2018oyd,Kanzi:2019gtu,Gullu:2020qzu,Oliveira:2021abg,Kanzi:2021cbg,Ovgun:2018ran,Sakalli:2023pgn,Mangut:2023oxa,Liu:2024lve,Liu:2024oas}. 

Meanwhile,~quasinormal modes~(QNMs) represent a cornerstone in the study of dynamical systems in relativistic astrophysics and gravitational wave physics. Characterized by complex eigenfrequencies,~where the real part corresponds to oscillatory behavior and the imaginary part governs exponential decay~\cite{Berti:2009kk, Sekhmani:2025zen, RibesMetidieri:2025lxr, Gu:2025lyz, Cao:2024oud},~QNMs emerge as transient solutions to perturbation equations in curved spacetime. These modes are most prominently observed during the ringdown process of black hole mergers~\cite{Berti:2009kk, Berti:2007dg, Zhao:2022lrl, Julie:2024fwy, Ma:2022wpv}, where the remnant black hole settles into equilibrium by gravitational wave emission~\cite{Lin:2022huh, Lai:2025nyo}. Their spectral signatures, detectable by instruments like LIGO~\cite{Fiziev:2019ewy, Bian:2021ini, Fan:2018vgw, Gao:2018tva, Chianese:2024nyw, Gardner:2024frf, Datta:2023vbs, Chan:2023atg, Suresh:2021rsn, Ashoorioon:2008nh, VanDenBroeck:2004wj}, Virgo~\cite{Ashoorioon:2008nh, VanDenBroeck:2004wj, Field:2025isp, Hongsheng:2018ibg, Babichev:2025ric}, and next-generation observatories such as the Einstein Telescope, encode critical information about spacetime geometry: black hole parameters (mass, spin and charge), and potential deviations from General Relativity.~Recent development in numerical relativity~\cite{Campanelli:2005dd, Baker:2005vv, Pretorius:2005gq, Roussille:2023sdr, Priyadarshinee:2023exb, Zhao:2023jiz, Chrysostomou:2022evl, Xing:2022emg, Daghigh:2022uws} and holography~\cite{BarraganAmado:2021uyw, Stanzer:2020ems, Pantelidou:2022ftm, Witczak-Krempa:2013xlz} (notably the AdS/CFT correspondence) have significantly broadened the theoretical applications of the QNMs. They now serve as probes for quantum gravity effects in near horizon~\cite{Poulias:2025eck}, hydrodynamic transport in quark-gluon plasmas~\cite{Kuntz:2025zde, Kuntz:2022kcw}, and even analog gravity systems in condensed matter physics~\cite{Sharma:2025hbk, Wang:2025lxr}.~Moreover,~the scalar field,~electromagnetic field and Dirac field perturbations for some solutions of the bumblebee model with a cosmological constant have also been studied~\cite{Singh:2025hor}. 

In Ref.~\cite{Liu:2024axg},~the authors elucidated the solutions of black holes in various aspects in the bumblebee theory.~In this paper, we will focus on a specific charged spherically symmetric black hole solution in the absence of a cosmological constant. The entire article will be constructed as follows: In Sec. \ref{II}, we will briefly review the bumblebee gravity theory and outline the derivation of the charged spherically symmetric black hole solution within this framework. In Sec. \ref{III}, we will derive the equations of motion for the scalar perturbation and the gravito-electromagnetic coupled perturbation. In Sec. \ref{IV}, we will introduce the continued fraction method~(CFM) and the asymptotic iteration method~(AIM), and simultaneously obtain the QNMs of two types of perturbations. Finally, we will present a summary in Sec. \ref{V}.

\section{Review of the Bumblebee Gravity Background}\label{II}
As discussed in the preceding section, the bumblebee model extends General Relativity by introducing a vector field, known as the bumblebee field, which couples nonminimally to gravity. In this section, we review the spherically symmetric black hole solution in bumblebee theory. The action can be read as~\cite{Liu:2024axg, Kostelecky:2003fs}
\begin{equation}\label{action}
\begin{aligned}
 S&=\int d^4 x\sqrt{-g}\left[\frac{1}{2\kappa}(R-2\Lambda)+\frac{\xi}{2\kappa} B^\mu B^\nu R_{\mu\nu}-\frac{1}{4} B_{\mu\nu} B^{\mu\nu}-V\left(B_\mu B^\mu\pm b^2\right)\right]\\&+\int d^4 x\sqrt{-g}\mathfrak{L}_M,
 \end{aligned}
 \end{equation}
where $\Lambda$ and $\kappa=\frac{8\pi G}{c^{4}}$ represent the cosmological constant and the gravitational constant, respectively. The potential $V$ is introduced to induce a nonvanishing vaccum expectation value (VEV) for the vector field $B_\mu$, and it is typically taken to have the functional form $V(B^\mu B_\mu \pm b^2)$. This implies that the VEV satisfies the constraint $B^\mu B_\mu = \pm b^2$. Accordingly, the vector field $B_\mu$ acquires a nontrivial VEV, denoted as $\langle B_\mu \rangle = b_\mu$, where $b_\mu$ is a spacetime-dependent vector satisfying $b^\mu b_\mu = \pm b^2 = \text{const}$. For notational convenience, we define the auxiliary variable $X \equiv B^\mu B_\mu \pm b^2$, which will be utilized in subsequent discussions. The anti-symmetric bumblebee field and electromagnetic field are given by 
\begin{eqnarray}
B_{\mu\nu}&=&\partial_\mu B_\nu-\partial_\nu B_\mu,\label{B}\\
F_{\mu\nu}&=&\partial_\mu A_\nu-\partial_\nu A_\mu.\label{F}
\end{eqnarray}
The electromagnetic field is nonminimally coupled to the bumblebee vector field, and the corresponding Lagrangian density $\mathfrak{L}_{M}$ is given by~\cite{Liu:2024axg}
\begin{equation}\label{LM}
\mathfrak{L}_M=\frac{1}{2\kappa}\left[F^{\alpha\beta} F_{\alpha\beta}+\gamma B^\mu B_\mu F^{\alpha\beta} F_{\alpha\beta}\right].
\end{equation}
Here, the parameter $\gamma$ represents the coupling between the electromagnetic field and the bumblebee field. The motion equations of the gravitational field, electromagnetic field and the bumblebee field can be obtained through variation with respect to $g_{\mu\nu}$,~$A_{\mu}$ and $B_{\mu}$,~respectively,
\begin{eqnarray}
 G_{\mu\nu}+\Lambda g_{\mu\nu}-\kappa T_{\mu\nu}^B-\kappa T_{\mu\nu}^M&=&0,\label{Einstein eq}\\
\nabla_{\mu}\left(F^{\mu\nu}+\gamma B^{\alpha}B_{\alpha}F^{\mu\nu}\right)&=&0,\label{Maxwell eq}\\
\nabla_\mu B^{\mu\nu}-2\left(\frac{\partial V(X)}{\partial X}B^\nu-\frac{\xi}{2\kappa}B_\mu R^{\mu\nu}-\frac{\gamma}{2\kappa}B^\nu F^{\alpha\beta}F_{\alpha\beta}\right)&=&0.\label{bumblebee eq}
\end{eqnarray}
Here,~the energy-momentum tensors of the bumblebee field and the bumblebee-coupled electromagnetic field take the form of
\begin{eqnarray}
&T_{\mu\nu}^B&=\frac{\xi}{\kappa}\left[\frac{1}{2} B^\alpha B^\beta R_{\alpha\beta} g_{\mu\nu}-B_\mu B^\alpha R_{\alpha\nu}-B_\nu B^\alpha R_{\alpha\mu}+\frac{1}{2}\nabla_\alpha\nabla_\mu\left(B_\nu B^\alpha\right)+\frac{1}{2}\nabla_\alpha\nabla_\nu\left(B_\mu B^\alpha\right)\right. \nonumber \\
&&~\left.-\frac{1}{2}\nabla^2\left(B_\mu B_\nu\right)-\frac{1}{2} g_{\mu\nu}\nabla_\alpha\nabla_\beta\left(B^\alpha B^\beta\right)\right]+2\frac{\partial V(X)}{\partial X} B_\mu B_\nu+B_\mu^{~\alpha} B_{\nu\alpha}\\
&&~-g_{\mu\nu}\left(V(X)+\frac{1}{4} B_{\alpha\beta} B^{\alpha\beta}\right),\label{TB} \\
&T_{\mu\nu}^M&=\frac{1}{\kappa}\left[\left(1+\gamma b^2\right)\left(2 F_{\mu\alpha} F_\nu^{~\alpha}-\frac{1}{2} g_{\mu\nu} F_{\alpha\beta} F^{\alpha\beta}\right)+\gamma B_\mu B_\nu F_{\alpha\beta} F^{\alpha\beta}\right].\label{TM} 
\end{eqnarray}
We consider the following metric ansatz describing a static and spherically symmetric spacetime: 
\begin{equation}\label{ds} 
 d s^2=-A(r) d t^{\prime2}+S(r) d r^2+r^2\left(d\theta^2+\sin^2\theta d\phi^2\right).
 \end{equation}
Because of the spherically symmetry,~we consider the electromagnetic 4-potential as
\begin{equation}\label{A}
 A_\mu \mathrm{d}x^\mu=\phi(r)\mathrm{d}t^{\prime}.
\end{equation}
Similar to Ref.~\cite{Casana:2017jkc},~we consider a spacelike background vector $b_\mu$ of the form 
\begin{equation}\label{b}
b_\mu=\left(0,b_r(r),0,0\right).
\end{equation}
By using the condition $b^\mu b_\mu=b^2=\text{const}$,~one can derive
\begin{equation}\label{br}
b_r(r)=b\sqrt{S(r)}.
\end{equation}
The analytic solution can be obtained by choosing the coupling parameter as $\gamma = \frac{\xi}{2 + l}$ and setting $l = \xi b^2$. By combining the ansatz~\eqref{ds}--\eqref{br} with the field equations~\eqref{Einstein eq}--\eqref{Maxwell eq}, and considering a quadratic potential of the form $V(X)=\frac{\lambda}{2}X^2$, one can solve the field equations and obtain a charged spherically symmetric black hole solution~\cite{Liu:2024axg}:
\begin{eqnarray}
 A(r)&=&1-\frac{2 M}{r}+\frac{2(1+l) Q_{0}^{2}}{(2+l) r^{2}},\label{Ar}\\
S(r)&=&\frac{1+l}{A(r)},\label{Sr}\\ 
\phi(r)&=&\sqrt{1+l}\frac{Q_0}{r}.\label{phi} 
\end{eqnarray}
With the above choice of $Q_0$,~the Maxwell invariant $F^{\mu\nu}F_{\mu\nu}=\frac{Q_0^2}{r^4}$ is independent of the Lorentz-violating parameter $l$.~Here, we introduce a new time variable $t=\sqrt{1+l}~t^{\prime}$~to make the form of the metric more symmetric, hence
\begin{equation}\label{newds} 
 d s^2=-f(r) d t^2+\frac{1}{f(r)}d r^2+r^2\left(d\theta^2+\sin^2\theta d\phi^2\right),
 \end{equation}
where $f(r)\equiv \frac{1}{S(r)}$.~According to this redefinition,~the bumblebee 4-potential and the electromagnetic 4-potential should be rewritten as
\begin{eqnarray}
b_r(r)&=&b\sqrt{\frac{1}{f(r)}},\\
A_\mu&=&\left(\frac{\phi(r)}{\sqrt{1+l}},0,0,0\right).
\end{eqnarray}
Similar to the Reissner-Nordström (RN) black hole,~this solution exhibits two horizons:
 \begin{eqnarray}
r_{1}&=&M\left(1-\sqrt{1-\frac{2(1+l) Q_{0}^{2}}{(2+l) M^{2}}}\right),\label{r1}\\ 
r_{2}&=&M\left(1+\sqrt{1-\frac{2(1+l) Q_{0}^{2}}{(2+l) M^{2}}}\right).\label{r2}
\end{eqnarray}

\section{The perturbation equations}\label{III}
In this section, we will derive the master equations of scalar, electromagnetic, and gravitational perturbations. Note that we treat these fields as test fields propagating on a fixed bumblebee-modified background.~This approach decouples their perturbation equations,~making the analysis tractable.~The self-consistent treatment of backreaction and bumblebee dynamics is reserved for future, more complex studies.
\subsection{Scalar field master equation}
We consider a free massless scalar field as the test field whose equation is the Klein-Gordon equation in non-flat spacetime
\begin{equation}\label{KG eq}
\partial_\mu(\sqrt{-g}g^{\mu\nu}\partial_\nu\Psi)=0.
\end{equation}
The spherically symmetry of the metric allows us to perform a spherical harmonic expansion on the field,
\begin{equation}\label{decompose}
\Psi(t,r,\theta,\phi)=\sum_{L=0}^{\infty}\sum_{m=-L}^{+L}\frac{\psi_{Lm}(r)}{r}Y_{Lm}(\theta,\phi)e^{-i\omega t}.
\end{equation}
Here,~$L$ and $m$ represent the orbital angular momentum quantum number and the magnetic quantum number, respectively. Recall the eigenequation of the spherical harmonics
\begin{equation}\label{spherical harmonics}
\frac{1}{\sin\theta}\frac{\partial}{\partial \theta}\left(\sin\theta\frac{\partial Y_{Lm}}{\partial \theta}\right)-\frac{m^2}{\sin^2\theta}Y_{Lm}+L(L+1)Y_{Lm}=0,
\end{equation}
\begin{figure}
    \centering
    \includegraphics[width=0.5\linewidth]{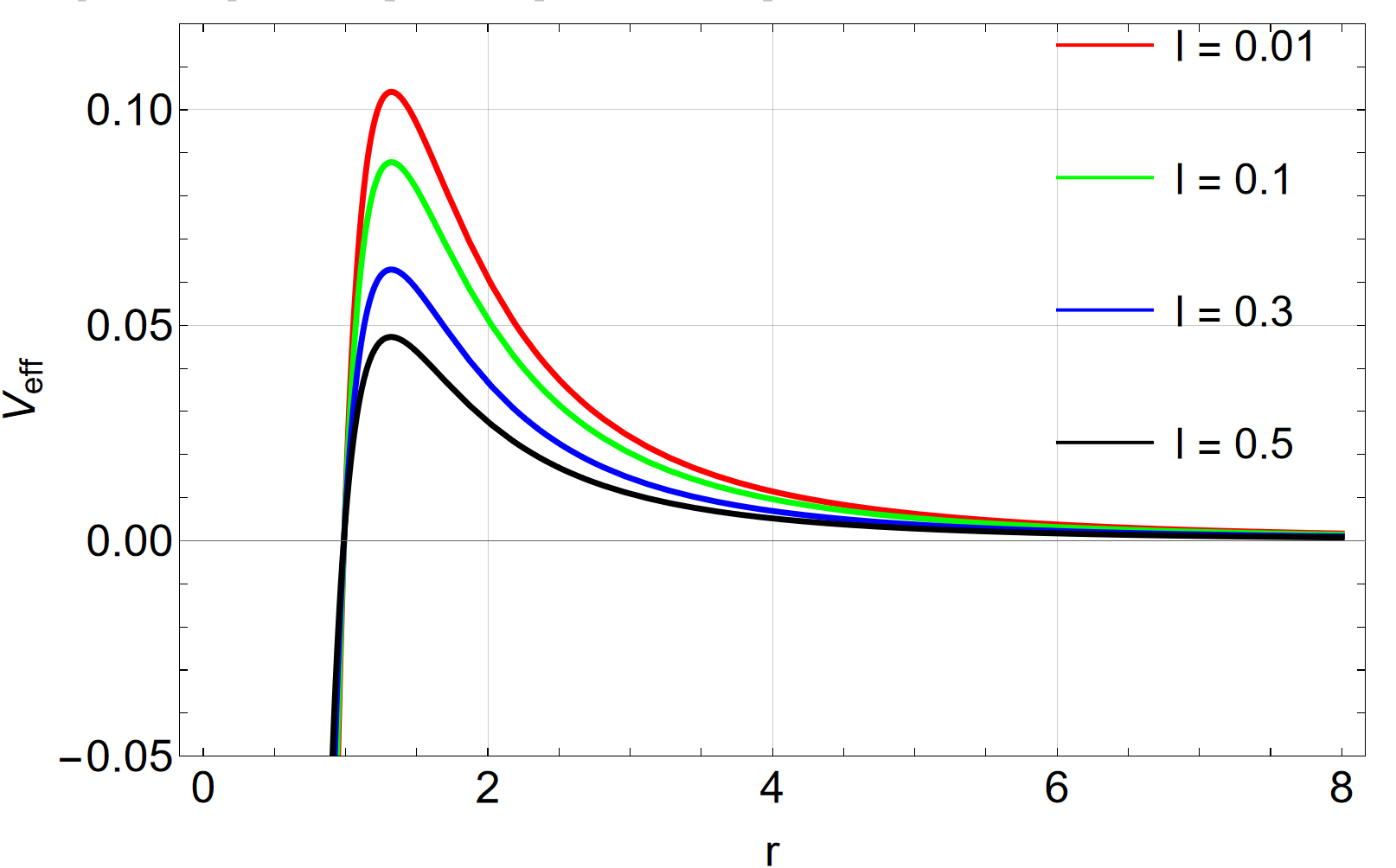}
    \caption{The effective potential of the scalar field with $Q_0=0.2M$ and varying Lorentz-violating parameter $l$.}
    \label{figure1}
\end{figure}
where we have replaced $\frac{\partial^2}{\partial \phi^2}$ with $-m^2$.~Substituting Eqs.~\eqref{decompose} and \eqref{spherical harmonics} into Eq.~\eqref{KG eq} with the specific form of the metric~\eqref{newds},~we can obtain the scalar perturbation master equation 
\begin{equation}\label{scalar master equation a} 
f(r)^{2}\frac{d^{2}\psi}{d r^{2}}+f(r)\frac{d f(r)}{d r}\frac{d\psi(r)}{d r}+\left(\omega^{2}-V_{\text{eff}}(r)\right)\psi(r)=0,
\end{equation}
where the effective potential read as
\begin{equation}\label{scalar potential}
V_{\text{eff}}(r)=f(r)\left(\frac{L(L+1)}{r^2}+\frac{1}{r}\frac{d f(r)}{d r}\right).
\end{equation}
Next,~we define the tortoise coordinate:
\begin{equation}\label{rstar}
r_{\star}=\int\frac{1}{f(r)} d r=(1+l)\left[r+\frac{1}{r_2-r_1}\left(r_2^2\ln\left(r-r_2\right)-r_1^2\ln\left(r-r_1\right)\right)\right].
\end{equation}
Here, we have already considered that the metric component $f(r)$ can be written as $f(r)=\frac{1}{1+l}\frac{\left(r-r_{1}\right)\left(r-r_{2}\right)}{r^{2}}$.~Therefore, the scalar perturbation equation can be rewritten as
\begin{equation}\label{scalar master equation b}
\left(\frac{d^2}{d r_\star^2}+\omega^2-V_\text{eff}(r)\right)\psi(r)=0 .
\end{equation}

\subsection{Gravito-electromagnetic field master equation}
 The perturbations of the gravitational field and electromagnetic field are given by
\begin{eqnarray}
A_{\mu}&=&A_{\mu}^{(0)}+a_{\mu},\label{perturbation A}\\
g_{\mu\nu}&=&g_{\mu\nu}^{(0)}+h_{\mu\nu},\label{perturbation g}
\end{eqnarray}
where $A_{\mu}^{(0)},~g_{\mu\nu}^{(0)}$ represent the background fields and $a_{\mu},~h_{\mu\nu}$ represent the vector and tensor perturbations, respectively. Additionally,~the components of the perturbation $h_{\mu\nu}$ can be classified into scalar, vector, and tensor types according to their transformation properties under spherical rotations.~Under spatial inversion, defined by the coordinate transformation $(\theta,\phi)\rightarrow(\pi-\theta,\pi+\phi)$, a quantity is said to have odd parity if it transforms with a factor of $(-1)^{L+1}$, or even parity if it transforms with $(-1)^{L}$. Therefore, the even and odd vector bases can be constructed by the spherical harmonics $Y_{L m}(\theta,\phi)$ as~\cite{Misner:1973prb, 9452501, Dewitt1973BlackH, Chandrasekhar1984, Regge:1957td}
\begin{eqnarray}
\left(V_{L m}^{1}\right)_{A}&=&\partial_{A} Y_{L m}(\theta,\phi),\label{V1}\\
\left(V_{L m}^{2}\right)_{A}&=&\varepsilon_{A C}\gamma^{B C}\partial_{B} Y_{L m}(\theta,\phi).\label{V2}
\end{eqnarray}
The even tensor bases $T_{L m}^{1}, T_{L m}^{1}$ and odd tensor base $T_{L m}^{3}$ can also be constructed as
\begin{eqnarray}
\left(T_{L m}^{1}\right)_{A B}&=&\nabla_{A}\nabla_{B} Y_{L m}(\theta,\phi),\label{T1}\\ 
\left(T_{L m}^{2}\right)_{A B}&=&\frac{L(L+1)}{2}\gamma_{A B} Y_{L m}(\theta,\phi),\label{T2}\\ 
\left(T_{L m}^{3}\right)_{A B}&=&\frac{1}{2}\left(\nabla_{A}\left(V_{L m}^{2}\right)_{B}+\nabla_{B}\left(V_{L m}^{2}\right)_{A}\right).\label{T3} 
\end{eqnarray}
Here, the capital letters $A, B$ denote the angular components $\theta$ and $\phi$, $\nabla$ and $\gamma^{A B}$ are covariant derivative and induced metric on the topological two-sphere $\mathbb{S}^2 $, $\varepsilon$ is the 2-dimensional totally antisymmetric tensor. Note that, in general, only perturbations with the same parity couple to each other, so we can study the odd and even parity perturbations seperately. Next, assuming that the time evolution of the perturbations depends on the exponential form $e^{-i\omega t}$, we can replace all time derivative operators with $-i\omega$. Besides, note that the spherical harmonics depends on $\phi$ through the form of $Y_{L m}(\theta,\phi)\sim e^{i m\phi}$.~Here, the magnetic quantum number $m$ does not directly affect the final form of the equations~\cite{Regge:1957td}. So we can set it to zero without loss of generality. In this paper, we use the Regge-Wheeler gauge~\cite{Regge:1957td}. Therefore, we can write the odd parts of the perturbations $h_{\mu\nu}$ and $a_{\mu}$ as
\begin{eqnarray}
h_{\mu\nu}^{\text{odd}}&=&\sum_L e^{-i\omega t}\left(\begin{array}{cccc} 0 & 0 & 0 & h_0(r) \\ 0 & 0 & 0 & h_1(r) \\ 0 & 0 & 0 & 0 \\ \star & \star & 0 & 0 \end{array}\right)\sin\theta\frac{\partial Y_{L 0}(\theta)}{\partial\theta}, \label{hodd} \\
a_{\mu}^{\text{odd}}&=&\sum_{L} e^{-i\omega t}\left(\begin{array}{c} 0 \\ 0 \\ 0 \\ a(r) \end{array}\right)\sin\theta\frac{\partial Y_{L 0}(\theta)}{\partial\theta}.\label{aodd} 
\end{eqnarray}
Here, $\star$ denotes the symmetric component.~By substituting Eqs.~\eqref{hodd} and \eqref{aodd} into the field equations \eqref{Einstein eq}-\eqref{TM}, we can obtain the equations of the electromagnetic and gravitational perturbations. Furthermore,~we reintroduce two perturbation notations $\Psi_{\text{g}},\Psi_{\text{em}}$ to distinguish the perturbations of the gravitational field and the electromagnetic field, which satisfy the following conditions
\begin{eqnarray} 
 \Psi_{\text{g}}&=&\frac{f(r)}{r} h_{1}(r),\label{psig}\\
\Psi_{\text{em}}&=&a(r).\label{psiem} 
\end{eqnarray}
Then the master equations can be read as
\begin{eqnarray}
\left(\frac{d^{2}}{d r_{\star}^{2}}+\frac{\omega^{2}}{1+l}\right)\Psi_{\text{g}}-V_{12}\Psi_{\text{em}}&=&V_{11}\Psi_{\text{g}}, \label{master equation a}\\
\left(\frac{d^2}{d r_\star^2}+\frac{\omega^{2}}{1+l}\right)\Psi_{\text{em}}-V_{21}\Psi_\text{g}&=&V_{22}\Psi_{\text{em}}.\label{master equation b} 
\end{eqnarray}
Here, the coefficient matrix $V$ can be expressed as
\begin{equation}\label{gra-elec potential}  
 V\equiv \left(\begin{array}{cc} V_{11} & V_{12} \\ V_{21} & V_{22} \end{array}\right)=\left(\begin{array}{cc} \frac{f(r)\left(8(1+l)Q_0^2+(2+l)r\left(-6M+rL(L+1)\right)\right)}{(1+l)(2+l)r^4} & -\frac{8 i Q_0\omega f(r)}{(1+l)(2+l) r^3} \\ \frac{if(r)(L^2+L-2) Q_0 }{\omega r^3} & \frac{f(r)\left(8Q_0^2+(2+l)L\left(1+L\right)r^2\right)}{(2+l)r^4} \end{array}\right). 
 \end{equation}
Besides, by introducing the fields~$\hat{\Psi}_{\text{g}},\hat{\Psi}_{\text{em}}$,
\begin{equation}\label{linear transformation}  
\begin{aligned}
&\left(\begin{array}{c} \hat{\Psi}_\text{g} \\ \hat{\Psi}_{\text{em}} \end{array}\right)=\left(\begin{array}{ll} \frac{i(L^2+L-2)Q_0}{\omega} & \\ & \frac{8iQ_0}{(1+l)(2+l)} \end{array}\right)\binom{\Psi_\text{g}}{\Psi_{\text{em}}},
\end{aligned} 
\end{equation}
we can make the effective potential matrix frequency-independent.~Then Eqs.~\eqref{master equation a} and~\eqref{master equation b} become
\begin{equation}\label{master equation c}  
 \left(\frac{d^2}{d r_\star^2}+\frac{\omega^2}{1+l}\right)\left(\begin{array}{c} \hat{\Psi}_\text{g} \\ \hat{\Psi}_{\text{em}} \end{array}\right)=\hat{V}\left(\begin{array}{c} \hat{\Psi}_\text{g} \\ \hat{\Psi}_{\text{em}} \end{array}\right), 
 \end{equation}
where the new effective potential matrix is
\begin{equation}\label{potential b}  
 \hat{V}\equiv \left(\begin{array}{cc} \frac{f(r)\left(8(1+l)Q_0^2+(2+l)r\left(-6M+rL(L+1)\right)\right)}{(1+l)(2+l)r^4} & -f(r)\frac{i(L^2+L-2)Q_0}{ r^3} \\ \frac{f(r)}{ r^3}\frac{8iQ_0}{(1+l)(2+l)} & \frac{f(r)\left(8Q_0^2+(2+l)L\left(1+L\right)r^2\right)}{(2+l)r^4} \end{array}\right).
\end{equation}

\section{Calculation methods for Quasinormal modes}\label{IV}
\subsection{QNMs of scalar field solved by CFM}

 For the scalar field perturbation $\psi$ with undetermined boundary conditions, there are many solutions allowed by the differential equation~\eqref{scalar master equation b}. To solve this equation we need to figure out the asymptotic behavior of the scalar perturbation. Since we only focus on the perturbation behavior outside the black hole horizon, we should consider the boundary conditions at both the event horizon $\left(r=r_{2}\right)$ and infinity $\left(r\rightarrow\infty\right)$. Note that,~in the absense of quantum effects, the physically relevant boundary conditions are pure ingoing waves at the horizon and pure outgoing waves at infinity. Besides, we only consider pure outgoing waves at infinity. The asymptotic behaviors are
\begin{eqnarray}
 \psi(r)&\sim& e^{-i\omega r_{\star}}\sim\left(r-r_2\right)^{-i(1+l)\omega\frac{r_2^2}{r_2-r_1}},~~~~~~ ~r_{\star}\rightarrow r_2,\label{BC1}\\
~~~~~~~~\psi(r)&\sim& e^{i\omega r_{\star}}\sim r^{2Mi(1+l)\omega  } e^{i(1+l)\omega r},~~~~~~~~~~~~r_{\star}\rightarrow+\infty.\label{BC2} 
\end{eqnarray}
Next, we will use the CFM \cite{Leaver:1985ax} to solve the differential equation~\eqref{scalar master equation b}. According to the asymptotic conditions \eqref{BC1} and \eqref{BC2}, we can write the trial solution of the scalar field:
\begin{equation}\label{scalar asymptotic solution}
\psi(r)\sim  r^{2i(1+l)\omega M}e^{i(1+l)\omega r}\left(\frac{r-r_2}{r-r_1}\right)^{-i(1+l)\omega\frac{r_2^2}{r_2-r_1}}\sum_{n=0}^{+\infty} a_n\frac{\left(r-r_2\right)^n}{\left(r-r_1\right)^n}.
\end{equation}
By substituting the series expansion \eqref{scalar asymptotic solution} into Eq.~\eqref{scalar master equation b}, a six-term recurrence relation for the expansion coefficients $a_{n}$ can be obtained
\begin{eqnarray} 
\alpha_0a_1+\beta_0a_0&=&0,\label{ir1}\\ 
\alpha_1a_2+\beta_1a_1+\gamma_1a_0&=&0,\label{ir2}\\ 
\alpha_2a_3+\beta_2a_2+\gamma_2a_1+\theta_2a_0&=&0,\label{ir3}\\ 
\alpha_3a_4+\beta_3a_3+\gamma_3a_2+\theta_3a_1+\mu_3a_0&=&0,\label{ir4}\\ 
\alpha_n a_{n+1}+\beta_n a_n+\gamma_n a_{n-1}+\theta_n a_{n-2}+\mu_n a_{n-3}+\nu_n a_{n-4}&=&0,~~~~n\ge4.\label{iteration relation}
\end{eqnarray}
Here, the explicit expressions of the coefficients in front of the $a_{i}$ are shown in the appendix \ref{A}. The iteration relation can be further expressed in matrix form
\begin{equation}\label{matrix 1}
\begin{pmatrix}
\beta_0 & \alpha_0 &        &     &   &        &        \\
\gamma_1 & \beta_1 & \alpha_1 &   &     &        &        \\
\theta_2 & \gamma_2 & \beta_2 & \alpha_2 &        &     &   \\
\mu_3 & \theta_3 & \gamma_3 & \beta_3 & \alpha_3 &      &  \\
\nu_4 & \mu_4 & \theta_4 & \gamma_4 & \beta_4 & \alpha_4 &\\
\vdots &        &     &    &         &         & \ddots
\end{pmatrix}
\begin{pmatrix}
a_{0} \\ a_{1} \\ a_{2} \\ a_{3} \\ a_{4} \\ \vdots
\end{pmatrix}
= 0. 
\end{equation}
In order for the scalar field to have solutions, the determinant of the coefficient matrix must vanish,
\begin{equation}\label{determination}
\left|
\begin{array}{ccccccc}
\beta_0 & \alpha_0 &        &        &        &      &  \\
\gamma_1 & \beta_1 & \alpha_1 &        &        &    &    \\
\theta_2 & \gamma_2 & \beta_2 & \alpha_2 &        &  &      \\
\mu_3 & \theta_3 & \gamma_3 & \beta_3 & \alpha_3 &   &     \\
\nu_4 & \mu_4 & \theta_4 & \gamma_4 & \beta_4 & \alpha_4 &\\
\vdots &         &         &         &     &    & \ddots
\end{array}
\right| = 0. 
\end{equation}
The final calculation results are listed in Table \ref{table1}.
\subsection{QNMs of scalar field solved by AIM}

\begin{figure}
    \centering
    \includegraphics[width=0.45\linewidth]{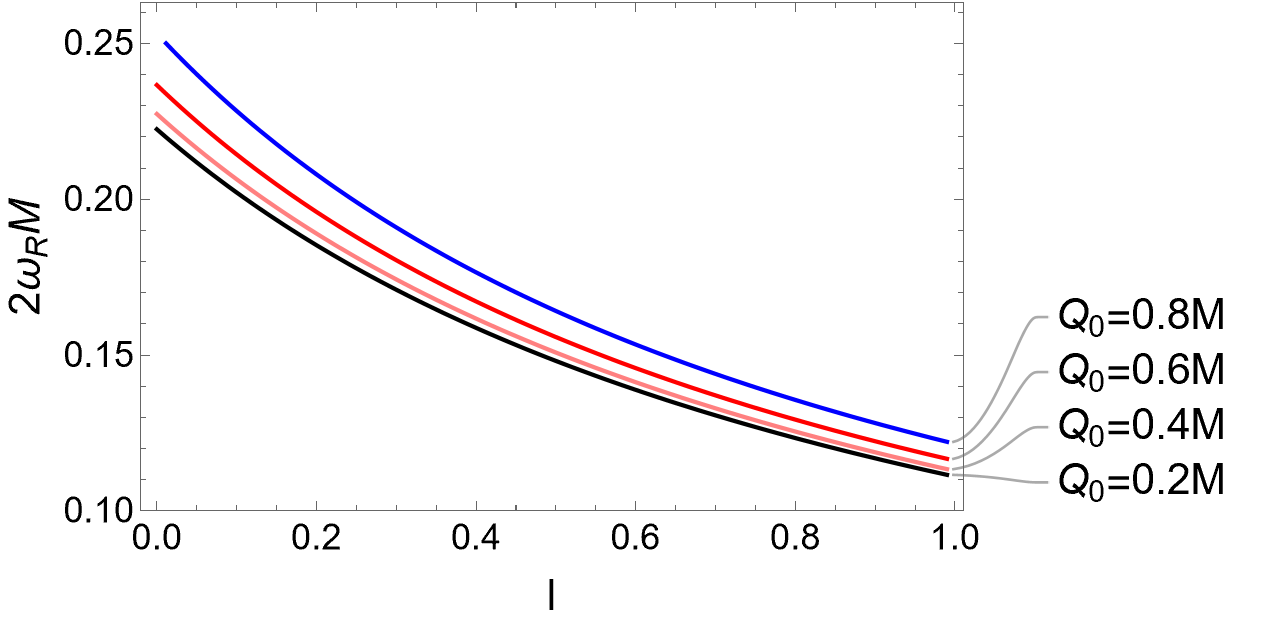}\hspace{0.05\linewidth}
\includegraphics[width=0.45\linewidth]{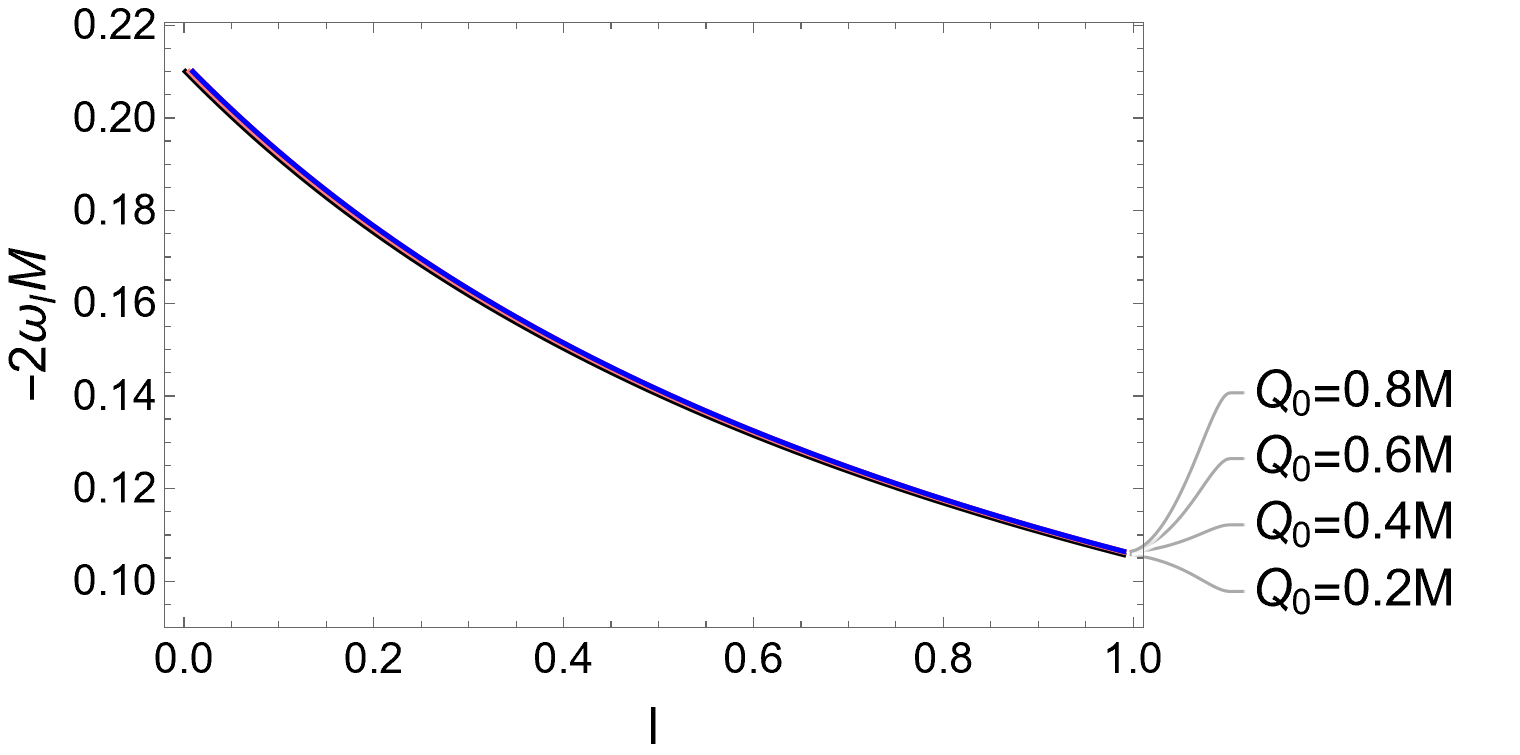}
    \caption{The real parts (left) and imaginary parts (right) of the QNFs of the scalar field by using the CFM for the black hole with varying electric charge $Q_0$ and Lorentz-violating parameter $l$ (ranging from $0$ to $1$).}
    \label{figure2}
    \includegraphics[width=0.45\linewidth]{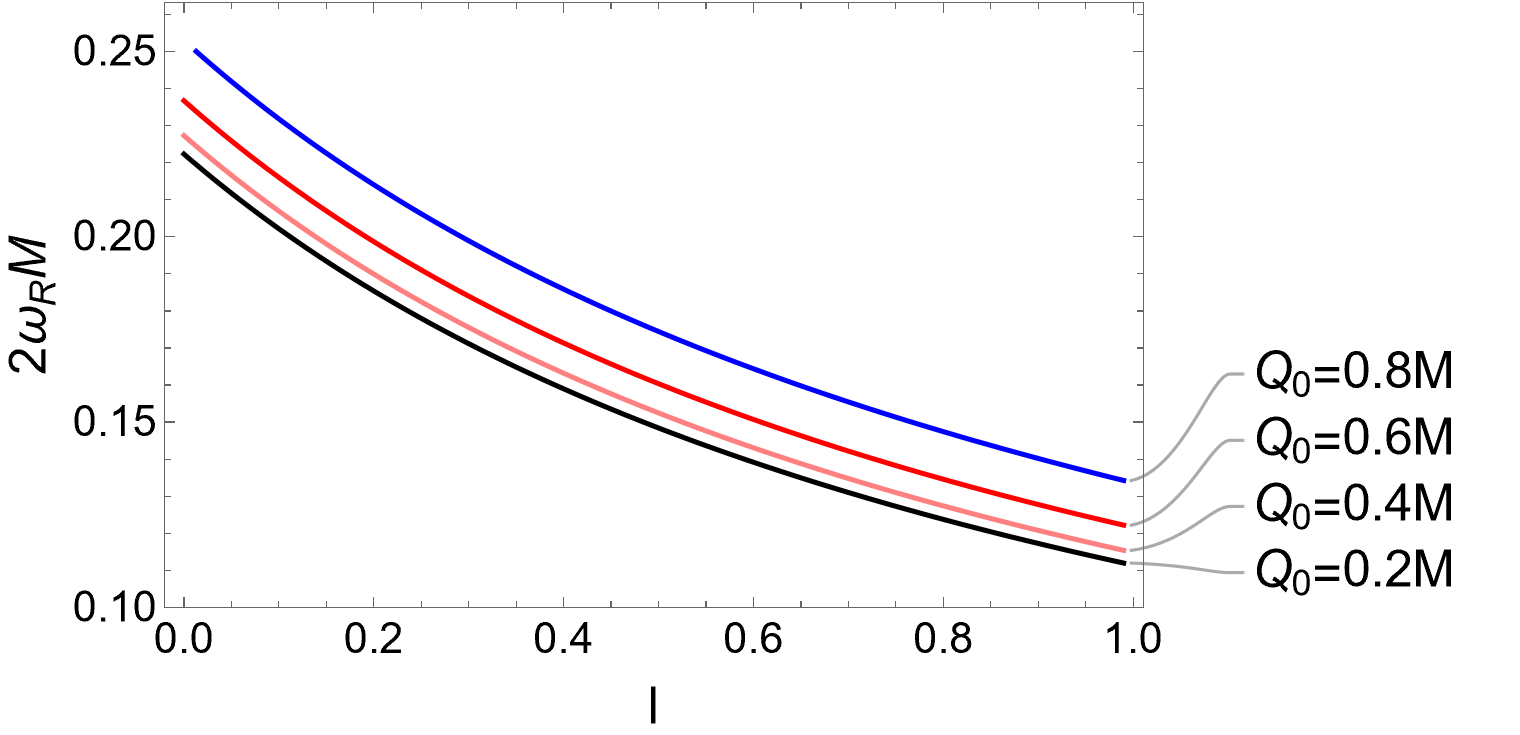}\hspace{0.05\linewidth}
\includegraphics[width=0.45\linewidth]{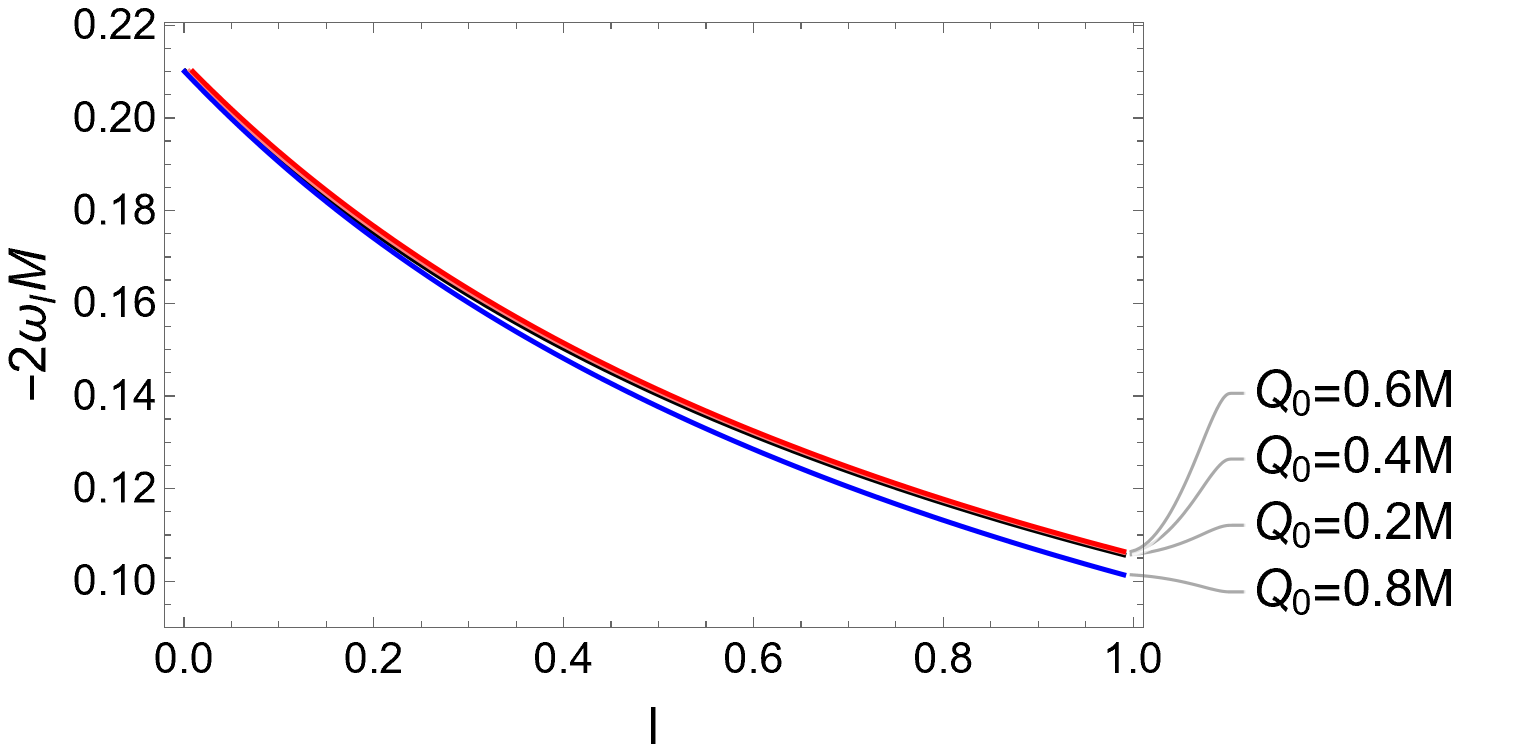}
    \caption{The real parts (left) and imaginary parts (right) of the QNFs of the scalar field by using the AIM for the black hole with varying electric charge $Q_0$ and Lorentz-violating parameter $l$ (ranging from $0$ to $1$).}
    \label{figure3}

\end{figure}

In this section we use the AIM~\cite{Saad:2003vhv, Cho:2011sf} to get the QNMs of the scalar field.~Its essence is actually a method for solving second-order differential equations from an asymptotic point of view.~Let us first consider a very conventional second-order ordinary differential equation
\begin{equation} \label{ODE1}
    \chi^{\prime\prime}(x)=\lambda_{0}(x)\chi^{\prime}(x)+s_{0}(x)\chi(x). 
\end{equation}
By taking the derivative of the above equation,~we can obtain
\begin{equation} \label{3rd}
\chi^{\prime\prime\prime}(x)=\left(\lambda_0^{\prime}+s_0+\lambda_0^2\right)\chi^{\prime}(x)+\left(s_0^{\prime}+s_0\lambda_0\right)\chi(x)=\lambda_1\chi^{\prime}(x)+s_1\chi(x), 
\end{equation}
where $\lambda_{1}(x)=\lambda_{0}^{\prime}(x)+s_{0}(x)+\lambda_{0}^{2}(x),~s_{1}(x)=s_{0}^{\prime}(x)+s_{0}(x)\lambda_{0}(x)$. Our task is to discern the pattern of coefficients that emerge after $n$ differentiations. Should the recurrence relation not become evident through this process, we will perform an additional differentiation of Eq.~\eqref{3rd} at this juncture
\begin{equation} \label{4th}
 \chi^{\prime\prime\prime\prime}(x)=\left(\lambda_1^{\prime}+s_1+\lambda_0\lambda_1\right)\chi^{\prime}(x)+\left(s_1^{\prime}+s_0\lambda_1\right)\chi(x)=\lambda_2\chi^{\prime}(x)+s_2\chi(x),
 \end{equation}
where $\lambda_{2}(x)=\lambda_{1}^{\prime}(x)+s_{1}(x)+\lambda_{0}(x)\lambda_{1}(x),~s_{2}(x)=s_{1}^{\prime}(x)+s_{0}(x)\lambda_{1}(x)$. In this way, we can summarize the recurrence relation of the coefficients after $n$ derivatives,
\begin{eqnarray} \label{n+2 th}
 \chi^{(n+2)}(x)&=&\lambda_{n}\chi^{\prime}(x)+s_{n}\chi(x),\nonumber \\ \lambda_{n}&=&\lambda_{n-1}^{\prime}+s_{n-1}+\lambda_{0}\lambda_{n-1}, \\ s_{n}&=&s'_{n-1}+s_{0}\lambda_{n-1}. \nonumber
\end{eqnarray}
For large $n$, an asymptotic relation holds:
\begin{equation} \label{delta_n}
 \delta_{n}=\left|\begin{array}{ll} \lambda_{n} & \lambda_{n-1} \\ s_{n} & s_{n-1} \end{array}\right|=0,
 \end{equation}
which implies that, as $n$ becomes sufficiently large,~the ratio of the coefficients tends toward a fixed function
\begin{equation}\label{fraction limitation}
  \frac{s_n(x)}{\lambda_n(x)}=\frac{s_{n-1}(x)}{\lambda_{n-1}(x)}=\beta(x).    
\end{equation}
Next, we consider its logarithmic derivative
\begin{equation}\label{logarithm}
\frac{d}{dx}\ln\chi^{(n + 1)}=\frac{\chi^{(n + 2)}}{\chi^{(n + 1)}}=\frac{\lambda_n\chi' + s_n\chi}{\lambda_{n - 1}\chi' + s_{n - 1}\chi}=\frac{\lambda_n(\chi'+\frac{s_n}{\lambda_n}\chi)}{\lambda_{n - 1}(\chi'+\frac{s_{n - 1}}{\lambda_{n - 1}}\chi)}\xrightarrow{n\text{ is large}}\frac{\lambda_n}{\lambda_{n - 1}}.
\end{equation}
Integrating the above relation yields
\begin{equation}\label{n+1 th}
\chi^{(n + 1)} = c_1e^{\int^x\frac{\lambda_n(t)}{\lambda_{n - 1}(t)}dt}=c_1e^{\int^x\frac{\lambda'_{n - 1}(t)+s_{n - 1}(t)+\lambda_0(t)\lambda_{n - 1}(t)}{\lambda_{n - 1}(t)}dt}=c_1\lambda_{n - 1}(x)e^{\int^x[\beta(t)+\lambda_0(t)]dt}.
\end{equation}
Therefore, we recall recurrence relation~\eqref{n+2 th} and obtain
\begin{equation}\label{h1}
\lambda_{n - 1}\chi' + s_{n - 1}\chi = c_1\lambda_{n-1} e^{\int^x[\beta(t)+\lambda_0(t)]dt}\equiv\lambda_{n - 1}f(x),
\end{equation}
where $f(x)$ is defined by $f(x)=c_1e^{\int^x[\beta(t)+\lambda_0(t)]dt} $. Dividing both sides by $\lambda_{n-1}$, we obtain a first-order nonhomogeneous ordinary differential equation
\begin{equation}\label{ODE2}
\chi'(x)+\beta(x)\chi(x)=f(x).
\end{equation}
It is evident that this differential equation can be
solved through the method of
variation of parameters. The
corresponding
general solution
reads:
\begin{equation}\label{Chi}
\chi(x)=e^{-\int^x\beta(t)dt}\left(c_2 + c_1\int^x e^{\int^t[2\beta(u)+\lambda_0(u)]du}dt\right).
\end{equation}
Note that we need to calculate $\lambda_n$ and $s_n$ in order to find $\beta(x)$. However, due to the derivative operations in the recurrence relation, the computational speed is affected to some extent. To solve this problem, we expand $\lambda_n$ and $s_n$ at 
$x=\xi$:
\begin{eqnarray}
\lambda_n(\xi)&=&\sum_{i = 0}^{\infty}c_n^i(x - \xi)^i,\label{lambda_n} \\
s_n(\xi)&=&\sum_{i = 0}^{\infty}d_n^i(x - \xi)^i. \label{s_n}
\end{eqnarray}
Substituting these expansions into the recurrence relation~\eqref{n+2 th},~we have
\begin{eqnarray}
c_n^i&=&(i + 1)c_{n + 1}^{i+1}+d_{n - 1}^i+\sum_{k = 0}^i c_0^k c_{n - 1}^{i - k},\label{c_n} \\
d_n^i&=&(i + 1)d_{n - 1}^{i + 1}+\sum_{k = 0}^i d_0^k c_{n - 1}^{i - k}.\label{d_n}
\end{eqnarray}
As $n$ is large enough, we obtain
\begin{equation}\label{h2}
d_n^0c_{n-1}^0-d_{n - 1}^0 c_n^0 = 0.
\end{equation}
Next, we use the above method to solve the master equation \eqref{scalar master equation b}. As a first step,~we perform a coordinate transformation 
\begin{equation}\label{Xi}
\xi=1-\frac{r_2}{r}
\end{equation}
to facilitate the analysis.~Note that we only focus on the area outside the black hole event horizon, so the range of $\xi$ is $(0,1)$. Here we need an additional step, that is to find the maximum point of the effective potential.~We can derive the explicit form of the effective potential through \eqref{scalar potential}:
\begin{equation}\label{scalar potential 2}
V_{\text{eff}}(r)=\frac{1}{1+l}\left(1-\frac{2 M}{r}+\frac{2(1+l) Q_{0}^{2}}{(2+l) r^{2}}\right)\left(\frac{L(L+1)}{r^2}+\frac{2M}{(1+l)r^3}-\frac{4 Q_0^2}{(2+l)r^4}\right).
\end{equation}
Then, employing the most straightforward derivative method yields the maximum point presented in Tab.~\ref{table1}. By substituting the coordinate transformation~\eqref{Xi} into the master equation \eqref{scalar master equation b}, we obtain
\begin{equation}\label{ODE xi}
\begin{aligned}
&\psi''(\xi)+\left(\frac{2}{\xi - 1}+\frac{r_2+(2\xi - 1)r_1}{[r_2+(\xi - 1)r_1]\xi}\right)\psi'(\xi)\\
&+\left({\frac{(1+l)^2r_2^4\omega^2}{[r_2+(\xi - 1)r_1]^2(\xi - 1)^4\xi^2}+\frac{r_1(2\xi^2 - 3\xi + 1)-r_2\left(\left(1+l\right)L(L+1) + 1-\xi\right)}{[r_2+(\xi - 1)r_1](\xi - 1)^2\xi}}\right)\psi(\xi)=0.
\end{aligned}
\end{equation}
After considering the boundary conditions, we adopt the following trial solution 
\begin{equation}
\psi(\xi)=(r - r_2)^{-i(1+l)\omega\frac{r_2^2}{r_2 - r_1}}r^{i(1+l)\omega\left(2M+\frac{r_2^2}{r_2 - r_1}\right)}e^{i(1+l)\omega r}\chi(\xi).
\end{equation}
Replacing $r$ with $\xi$, we can obtain the proportional form of the above trial solution:
\begin{equation}\label{scalar asymptotic solution 2}
\psi(\xi)\sim\xi^{-i(1 + l)\omega\frac{r_2^2}{r_2 - r_1}}(1 - \xi)^{-2i(1 + l)\omega M}e^{i(1 + l)\omega\frac{r_2}{1-\xi}}\chi(\xi).
\end{equation}
By substituting the trial condition~\eqref{scalar asymptotic solution 2} into Eq.~\eqref{ODE xi}, we obtain a second-order ordinary differential equation in the form of Eq. \eqref{ODE1}.~The specific expressions of the coefficients are not presented here due to the complexity of the expressions of the CFM and the AIM. The results calculated by the three methods~(CFM,~AIM,~and WKB~\cite{Iyer:1986np,Leaver:1990zz}) are comprehensively tabulated in Tab.~\ref{table1} and graphically represented in~Figs.~\ref{figure2}-\ref{figure3}.~It is evident that in the absence of
the Lorentz-violating effects (i.e. in the RN case), all three methods agree with each other very well,~thereby validating the reliability of
the computational approaches employed in this work.
\begin{table}[h]
\centering
\begin{tabular}{|c|c|c|c|c|}
\hline
 $Q_0/2M$&maximum point $r_{\text{max}}/2M$ &CFM&AIM&WKB \\ \hline
&&$2M\omega_R~~~-2M\omega_I$&~~$2M\omega_R~~-2M\omega_I$&~$2M\omega_R~~~-2M\omega_I$ \\ \hline
 0   &1.333333& 0.22096~~~~0.20974 & 0.22096~~0.20975 & 0.22093~~0.20974  \\
 0.1 &1.322419& 0.22253~~~~0.21008 & 0.22253~~0.21008 & 0.22250~~0.21004  \\
 0.2 &1.288653&  0.22755~~~~0.21097 & 0.22750~~0.21096 & 0.22745~~0.21120  \\
 0.3 &1.228502&  0.23698~~~~0.21192 & 0.23691~~0.21200 & 0.23708~~0.21271  \\
 0.4 &1.134313&  0.25299~~~~0.21016 & 0.25312~~0.21022 & 0.25409~~0.21337  \\
\hline
\end{tabular}
\caption{Fundamental quasinormal frequencies~(QNFs) of the scalar field for the charged black hole without Lorentz-violating effects,~i.e.~$l=0$,~using the CFM,~AIM,~WKB,~respectively. For different values of the electric charge $Q_0$,~an error of $+0.8\%$ in WKB is presented.}
\label{table1}
\end{table}

From Figs.~\ref{figure2}-\ref{figure3} we can see that in the case of the scalar perturbation, as $l$ gradually increases, the absolute values of both the real part and imaginary part of the QNMs gradually decrease.~As the electric charge increases,~the real part of the QNMs exhibits a noticeable growth,~while the imaginary part remains largely unaffected.~However, considering that the Lorentz-violating parameter is typically constrained to a small value, its influence on the QNMs is expected to be very small.

\subsection{QNMs of gravito-electromagnetic field solved by CFM}

Unlike the scalar field,~one problem that we need to face is that the gravitational-electromagnetic mode cannot be decoupled into two independent perturbation equations. It turns out that when we use the AIM, we cannot find a suitable recurrence relation.~There is a point that we need to emphasize. Once we let $l$ tends to zero, the coupling between gravitational and electromagnetic perturbations will disappear. This allows us to apply the AIM again to solve for the QNMs in the background without Lorentz-violating effects~(i.e. $l=0$), which will be reflected in the subsequent data in Tabs.~\ref{table2} and~\ref{table3}.  

Then we use the CFM to figure out the QNMs of the gravito-electromagnetic coupled perturbation equations.~As in previous analyses, we begin by
examining the asymptotic behaviors of the effective potential at the event horizon and at spatial infinity
\begin{eqnarray}   
\lim_{r\rightarrow r_{2}}\hat{V}&\rightarrow& 0, \quad \lim_{r\rightarrow\infty}\hat{V}\rightarrow0. 
\end{eqnarray}
And by solving the master equation \eqref{master equation c} both at the horizon and at infinity, we can obtain the asymptotic behaviors of the perturbations
\begin{eqnarray}
\lim_{r\rightarrow r_{2}}\hat{\Psi}_{\text{g}}&\rightarrow& e^{-i\frac{\omega}{\sqrt{1+l}}{r_\star}}, ~~\quad \lim_{r\rightarrow\infty}\hat{\Psi}_{\text{g}}\rightarrow e^{i\frac{\omega}{\sqrt{1+l}}{r_\star}}, \label{psig asymptotic}\\
\lim_{r\rightarrow r_{2}}\hat{\Psi}_{\text{em}}&\rightarrow& e^{-i\frac{\omega}{\sqrt{1+l}}{r_\star}}, ~\quad \lim_{r\rightarrow\infty}\tilde{\Psi}_{\text{em}}\rightarrow e^{i\frac{\omega}{\sqrt{1+l}}{r_\star}}.\label{psiem asymptotic}
\end{eqnarray}
With the boundary conditions \eqref{psig asymptotic} and \eqref{psiem asymptotic}, it turns out to be an eigenequation problem to solve the master equation \eqref{master equation c}.~And we can write the trial solution of the perturbations as before:
\begin{eqnarray}  \hat{\Psi}_{\text{g}}(r)&=&\left(\frac{r-r_2}{r-r_1}\right)^{-i\sqrt{1+l}~\omega\frac{r_{2}^{2}}{r_{2}-r_{1}}} r^{2i\sqrt{1+l}~\omega M } e^{i\sqrt{1+l}~\omega r}\sum_{n=0}^{+\infty}a_n^\text{g}\left(\frac{r-r_2}{r-r_1}\right)^{n},\label{psig asymptotic solution}\\ 
\hat{\Psi}_{\text{em}}(r)&=&\left(\frac{r-r_2}{r-r_1}\right)^{-i\sqrt{1+l}~\omega\frac{r_{2}^{2}}{r_{2}-r_{1}}} r^{2i\sqrt{1+l}~\omega M } e^{i\sqrt{1+l}~\omega r}\sum_{n=0}^{+\infty}a_n^{\text{em}}\left(\frac{r-r_2}{r-r_1}\right)^{n}\label{psiem asymptotic solution}.
\end{eqnarray}
By inserting the trial solutions~\eqref{psig asymptotic solution} and~\eqref{psiem asymptotic solution} into Eq.~\eqref{master equation c}, a sixth-term recurrence relation can be obtained:
\begin{eqnarray} 
\alpha_0\mathbf{A_1}+\beta_0\mathbf{A_0}&=&0,\label{IR1}\\ 
\alpha_1\mathbf{A_2}+\beta_1\mathbf{A_1}+\gamma_1\mathbf{A_0}&=&0,\label{IR2}\\ 
\alpha_2\mathbf{A_3}+\beta_2\mathbf{A_2}+\gamma_2\mathbf{A_1}+\theta_2\mathbf{A_0}&=&0,\label{IR3}\\ 
\alpha_3\mathbf{A_4}+\beta_3\mathbf{A_3}+\gamma_3\mathbf{A_2}+\theta_3\mathbf{A_1}+\mu_3\mathbf{A_0}&=&0,\label{IR4}\\ 
\alpha_n \mathbf{A_{n+1}}+\beta_n \mathbf{A_n}+\gamma_n \mathbf{A_{n-1}}+\theta_n \mathbf{A_{n-2}}+\mu_n \mathbf{A_{n-3}}+\nu_n \mathbf{A_{n-4}}&=&0,~~~~n\ge4.\label{Iteration Relation}
\end{eqnarray}
Here, \begin{equation}\mathbf{A_n}\equiv \binom{a_n^\text{g}}{a_n^{\text{em}}}.
\end{equation}
It should be noted that each coefficient is a $2\times2$ matrix,~for example
\begin{equation}
\alpha_n\equiv\begin{pmatrix}
\alpha_n^{11}&\alpha_n^{12} \\
\alpha_n^{21}&\alpha_n^{22}
\end{pmatrix}.
\end{equation}
As we did before, for the equation to have non-zero solutions, its determinant of the coefficient matrix must be zero, and the QNFs $\omega$ can be solved.

The computed results of the QNFs varying with the electric charge $Q_0$ are listed in Tabs.~\ref{table2}-\ref{table7}.~It is quite clear that when the Lorentz-violating parameter approaches to zero~(i.e. $l=0$), the metric smoothly reduces to the RN black hole metric. As expected, the QNFs of gravitational and electromagnetic perturbations reduce to that of the RN black hole.~For $L=2,3$,~the results from the three methods mutually verify each other and show good agreement.~However,~for $L=4$,~while the results from the CFM and the WKB are consistent with each other, the deviation of the AIM becomes more significant.~This occurs because,~at higher $L$ values, the effective potential
$$V_{\text{eff}}\propto \frac{L(L+1)}{r^2}$$becomes higher and sharper,~causing the solution of the differential equation to oscillate rapidly near the potential barrier. The AIM relies on expanding the coefficients around a specific point $\xi$ (Eqs. \eqref{lambda_n}-\eqref{s_n}),~and this expansion may fail to accurately capture the complex behavior of high-frequency modes,~thereby introducing numerical errors.~In contrast,~the CFM solves the frequency equation directly by expressing the solution as a series and deriving a recurrence relation (Eqs. \eqref{ir1}-\eqref{iteration relation} as well as Eqs. \eqref{IR1}-\eqref{Iteration Relation}).~The CFM is generally more stable for modes with higher angular momentum quantum number $L$,~as it is based on global convergence and is not influenced by the choice of a local expansion point.

Note that Eq.\eqref{master equation c} is a coupled equation. Physically, this coupling implies that dynamic perturbations near the event horizon undergo continuous mutual conversion and intertwining between gravitationally‑dominated and electromagnetically‑dominated components as they propagate outward. Consequently, the resulting QNMs describe the collective excitations of the entire coupled system, rather than purely independent gravitationally‑dominated or electromagnetically‑dominated oscillations. In this sense, each QNM frequency represents a mixed mode of the system.~To characterize the dominant nature of each mode, we follow the method introduced in Ref. \cite{Gu:2025lyz}. Specifically, we compute the modulus ratio $$R\equiv\frac{ \left | \Psi_g \right | }{\left | \Psi_{em} \right | }$$ of the eigenvector components under an suitably normalized quantity. For a given mode, $R\gg 1$, indicates that the perturbative energy is primarily concentrated in the gravitationally‑dominated component; we therefore identify it as a gravitationally‑dominated mixed mode. Conversely,~$R\ll 1$ signifies that the electromagnetic component dominates,~and the mode is classified as an electromagnetically‑dominated mixed mode.~The numerical values used to determine which branch the data belongs to are listed in the Tab. \ref{table8}.~As expected, the coupling strength between the electromagnetically‑dominated and gravitationally‑dominated parts increases with the Lorentz-violation parameter $l$.
\begin{table}[h]
\centering
\begin{tabular}{|c|c|c|c|c|c|}
\hline
 $Q_0/2M$ & CFM &  AIM & WKB \\
\hline
&$2M\omega_R~~~-2M\omega_I$&$2M\omega_R~~~-2M\omega_I$&$2M\omega_R~~~~-2M\omega_I$ \\ \hline
 0   &0.74734~~0.17793& 0.74734~~0.17792 & 0.74734~~0.17792 \\
 0.1 &0.74955~~0.17814& 0.74949~~0.17815 & 0.74938~~0.17801\\
 0.2 &0.75696~~0.17881& 0.75687~~0.17880 & 0.75687~~0.17880 \\
 0.3 &0.77256~~0.17973& 0.77244~~0.17963 & 0.77235~~0.17948\\
 0.4 &0.80244~~0.17985& 0.80244~~0.17929 & 0.80244~~0.17929 \\
\hline
\end{tabular}
\caption{The fundamental frequencies of the QNMs of the gravitationally‑dominated perturbation for the charged black hole without Lorentz-violating effects, i.e.~$l=0$ (using CFM,~AIM, and WKB), for various values of the charge $Q_0$. The angular momentum $L=2$, and the results are in good agreement with those obtained in Ref.~\cite{Guo:2022rms}.}
\label{table2}
\end{table}
\begin{table}[h]
\centering
\begin{tabular}{|c|c|c|c|c|c|}
\hline
 $Q_0/2M$ & CFM &  AIM &WKB \\
\hline
&$2M\omega_R~~~-2M\omega_I$&$2M\omega_R~~~-2M\omega_I$&$2M\omega_R~~~~-2M\omega_I$ \\ \hline
 0   &0.91519~~0.19001& 0.91519 0.19001 & 0.91519~~0.19001 \\
 0.1 &0.92593~~0.19073& 0.92593 0.19075 & 0.92593~~0.19076\\
 0.2 &0.95953~~0.19286& 0.95985 0.19288 & 0.95985~~0.19288 \\
 0.3 &1.02403~~0.19600& 1.02402 0.19603 & 1.02402~~0.19605\\
 0.4 &1.14032~~0.19808& 1.14026 0.19814 & 1.14026~~0.19814 \\
\hline
\end{tabular}
\caption{The fundamental frequencies of the QNMs of the electromagnetically‑dominated perturbation for the charged black hole without Lorentz-violating effects, i.e.~$l=0$ (using CFM,~AIM and WKB), for various values of the charge $Q_0$. The angular momentum $L=2$, and the results are in good agreement with those obtained in Ref.~\cite{Guo:2022rms}.}
\label{table3}
\end{table}

\begin{figure}
    \centering
\includegraphics[width=0.45\linewidth]{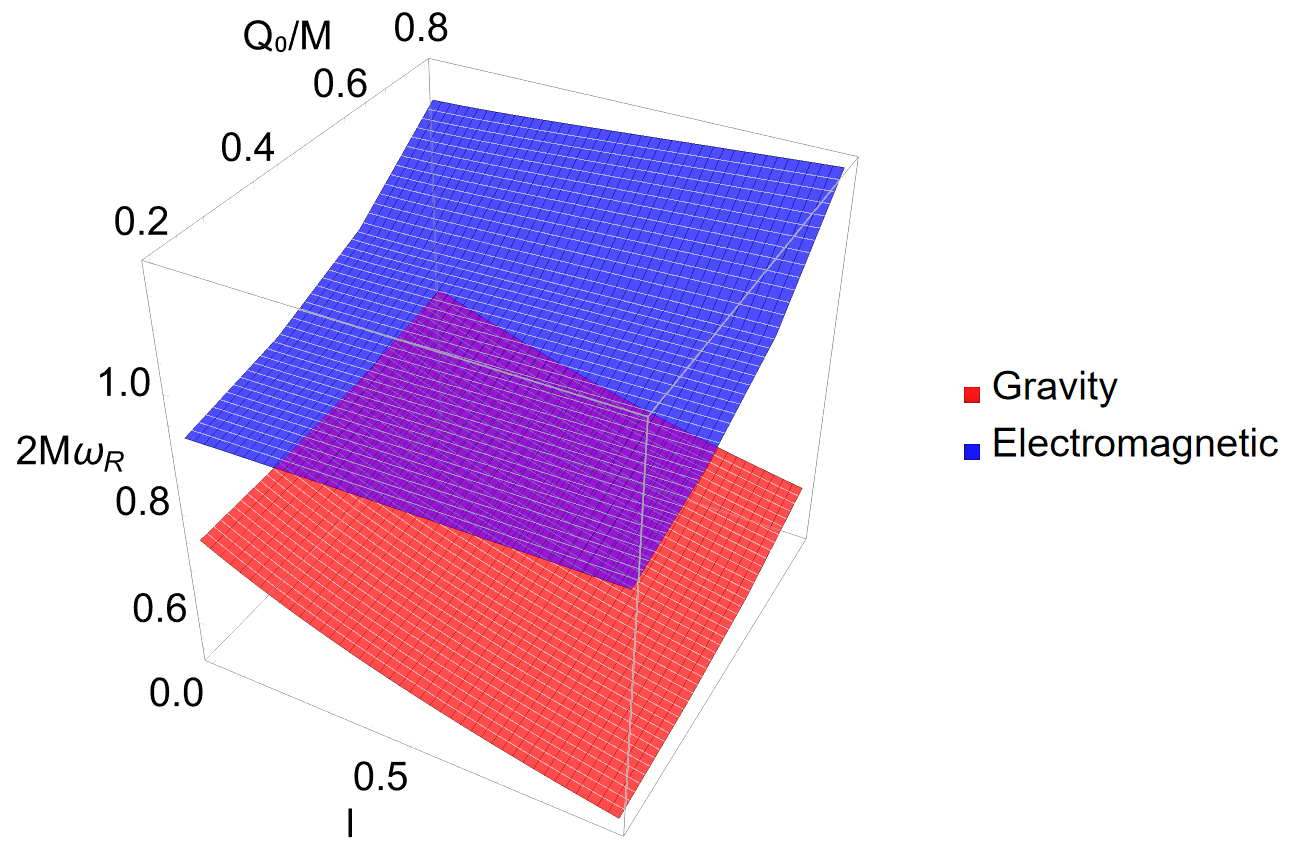}
\hspace{0.05\linewidth}\includegraphics[width=0.45\linewidth]{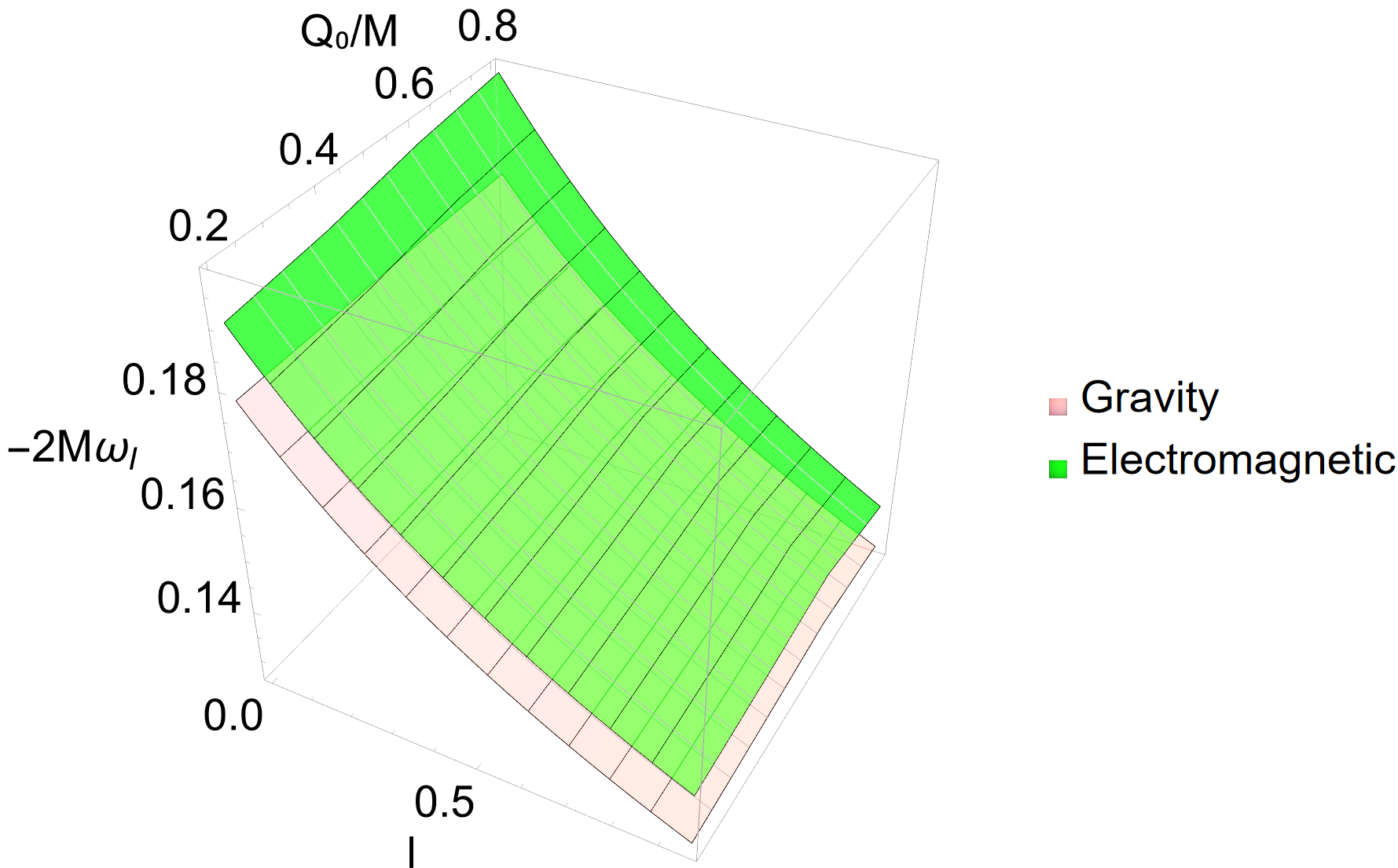}
    \caption{The real parts (left) and imaginary parts (right) of the QNFs of the gravitationally‑dominated perturbation and the electromagnetically‑dominated perturbation for the black hole with varying electric charge $Q_0$ (ranges from $0$ to $0.8M$) and Lorentz-violating parameter $l$ (ranges from $0$ to $1$),~and the angular momentum $L=2$.}
    \label{figure5}
\end{figure}

\begin{figure}
    \centering
\includegraphics[width=0.45\linewidth]{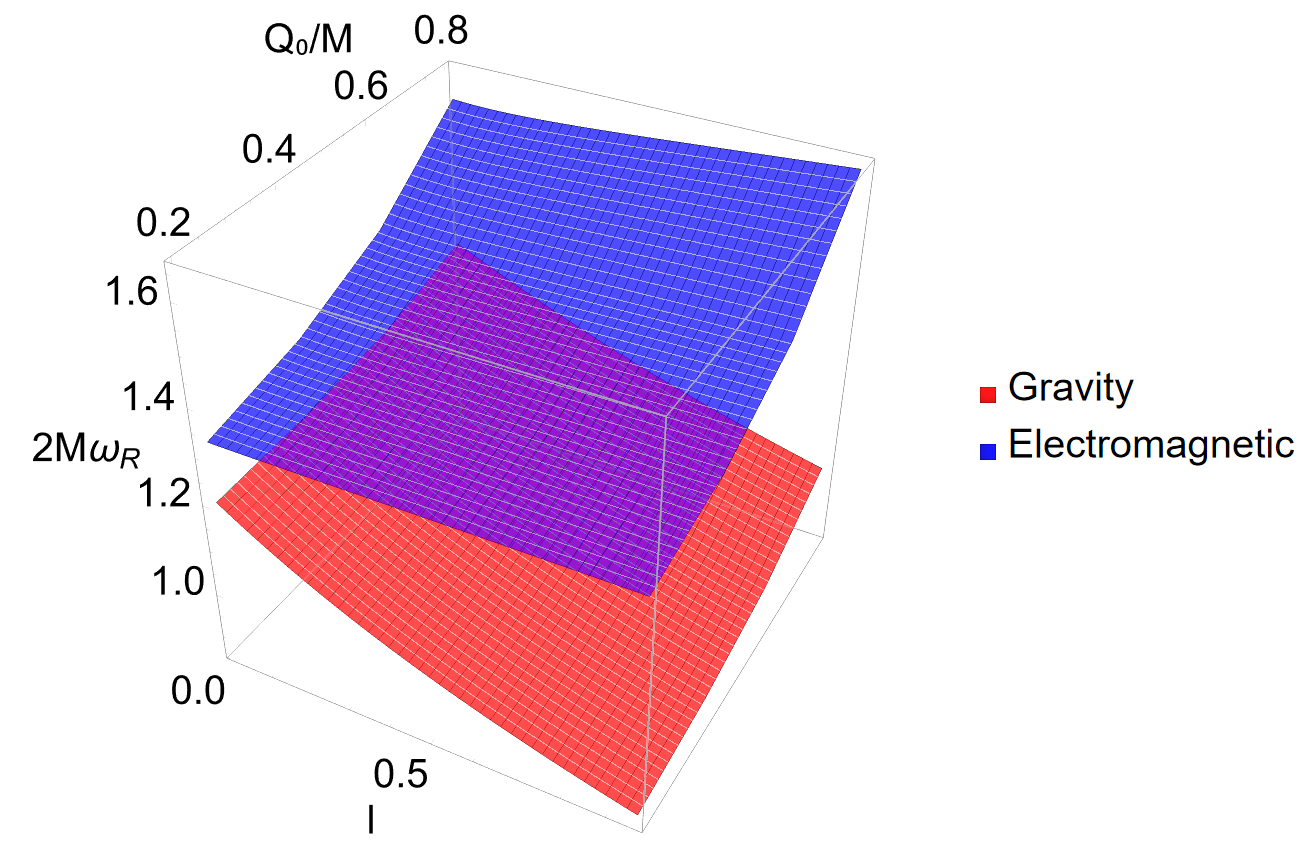}
\hspace{0.05\linewidth}\includegraphics[width=0.45\linewidth]{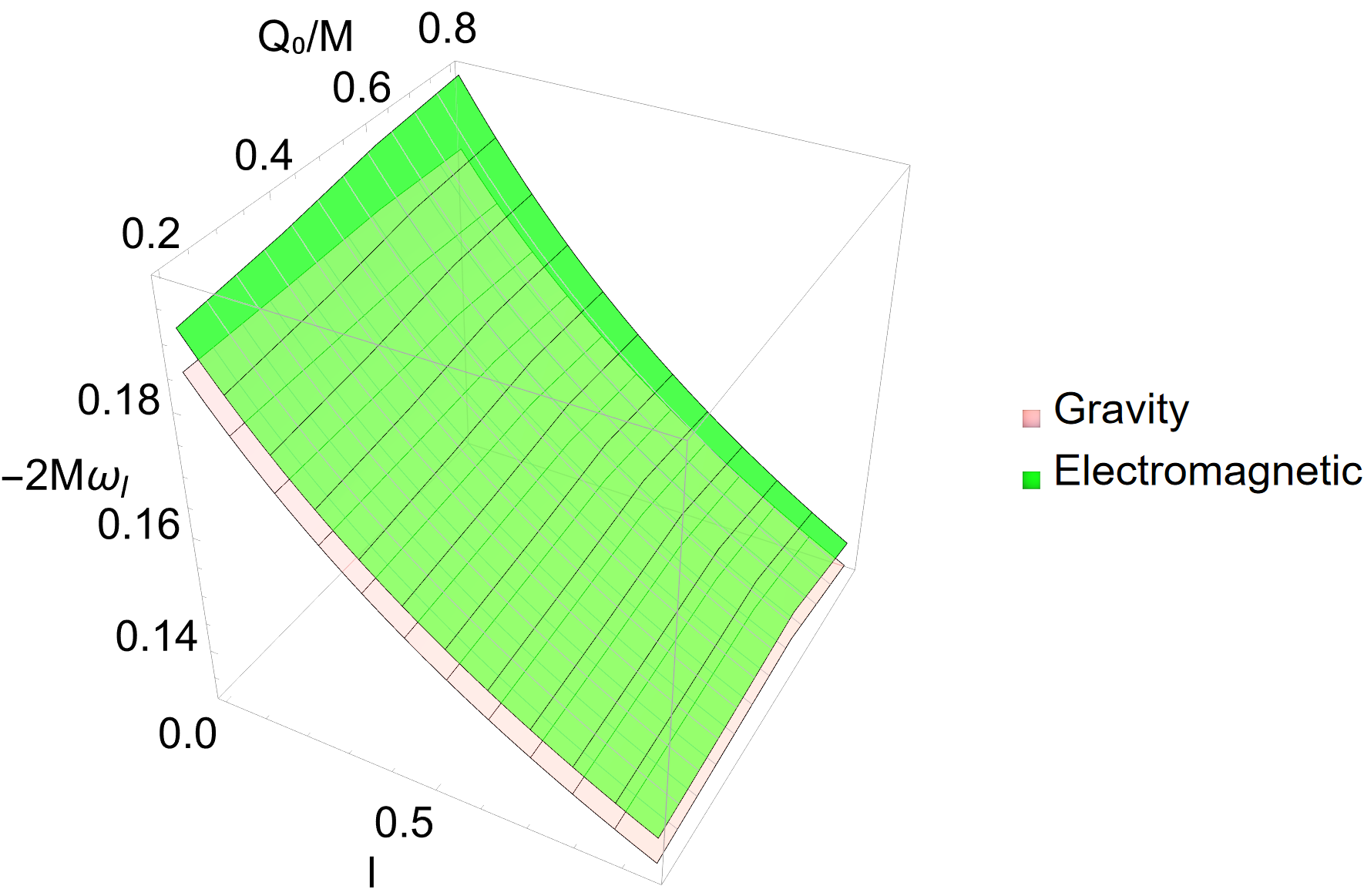}
    \caption{The real parts (left) and imaginary parts (right) of the QNFs of the gravitationally‑dominated perturbation and the electromagnetically‑dominated perturbation for the black hole with varying electric charge $Q_0$ (ranges from $0$ to $0.8M$) and Lorentz-violating parameter $l$ (ranges from $0$ to $1$),~and the angular momentum $L=3$.}
    \label{figure6}
\end{figure}

\begin{figure}
    \centering
\includegraphics[width=0.45\linewidth]{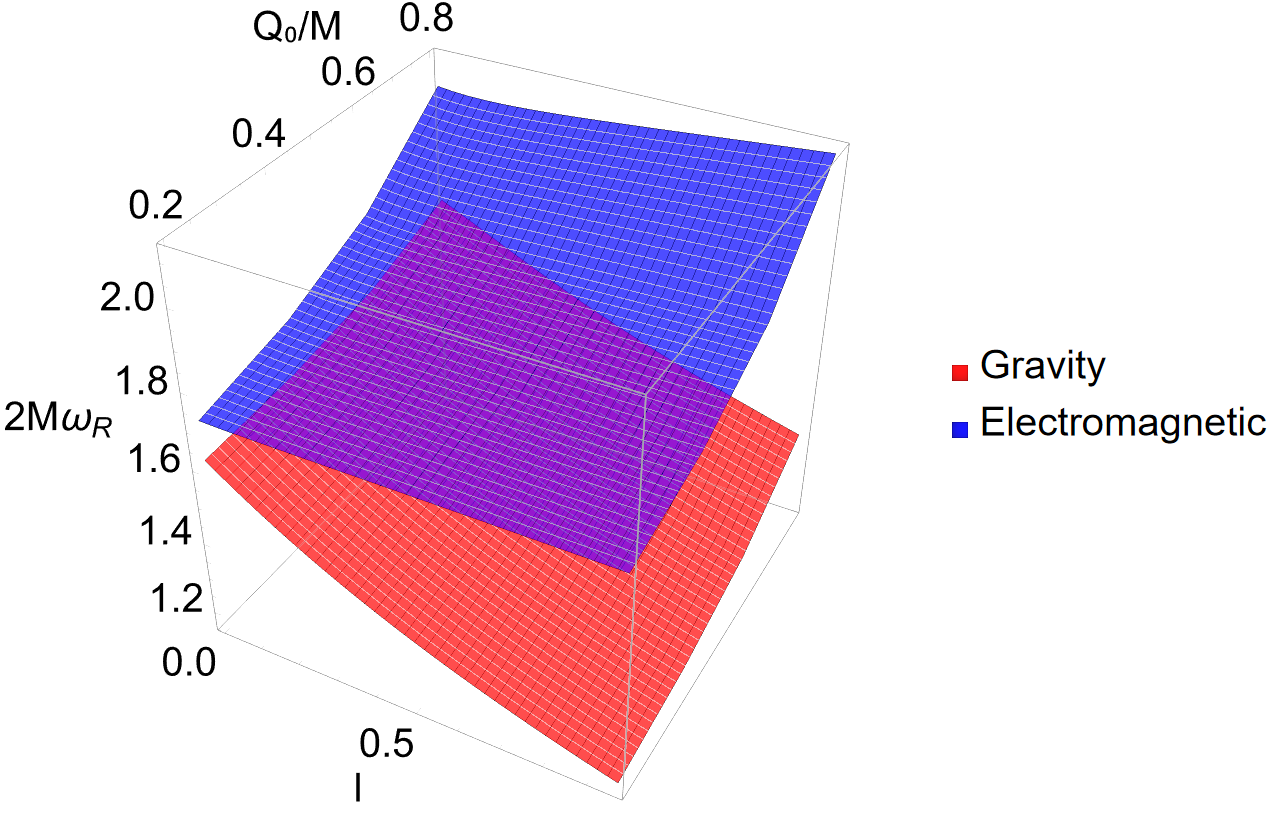}
\hspace{0.05\linewidth}\includegraphics[width=0.45\linewidth]{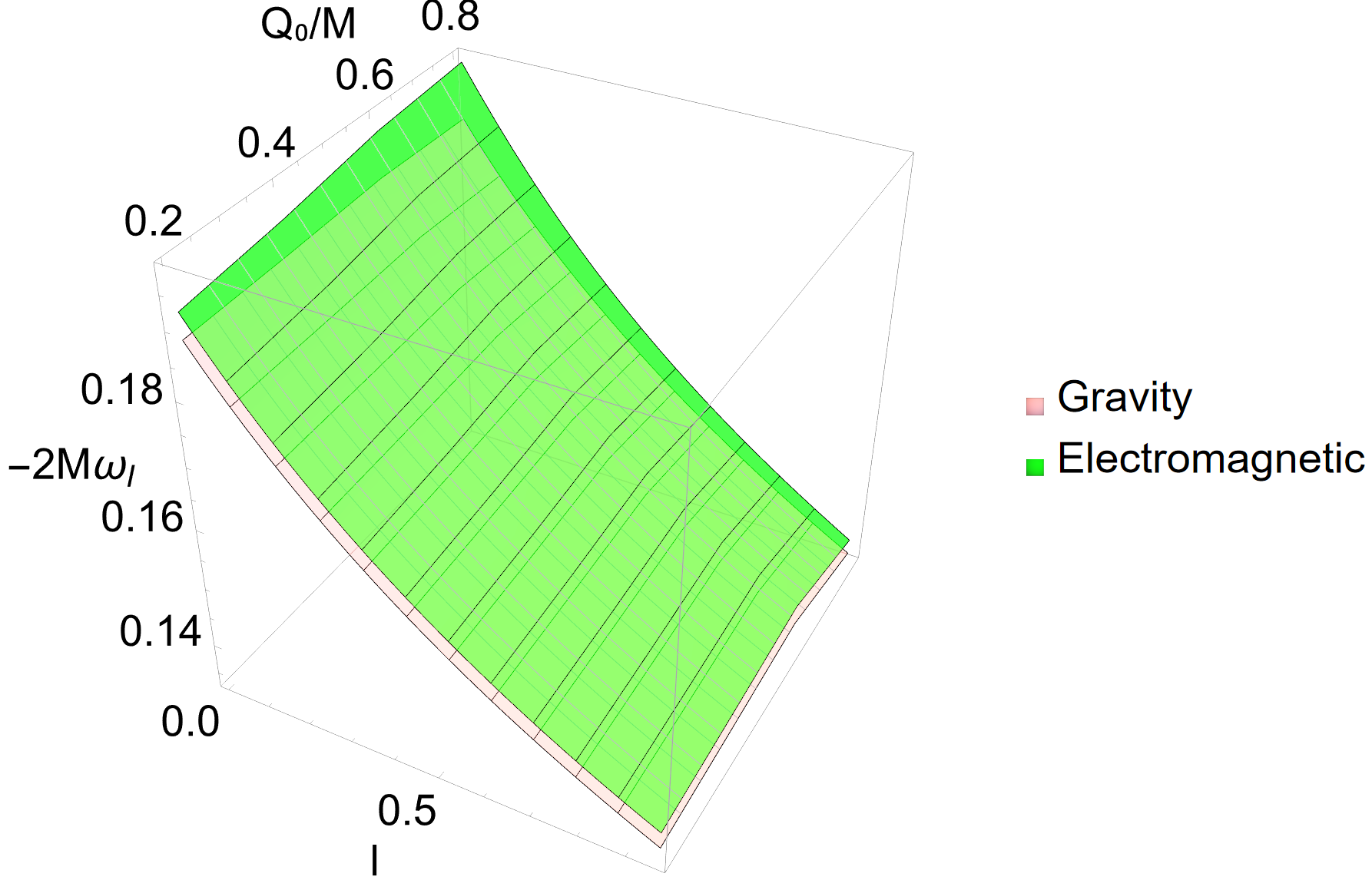}
    \caption{The real parts (left) and imaginary parts (right) of the QNFs of the gravitationally‑dominated perturbation and the electromagnetically‑dominated perturbation for the black hole with varying electric charge $Q_0$ (ranges from $0$ to $0.8M$) and Lorentz-violating parameter $l$ (ranges from $0$ to $1$),~and the angular momentum $L=4$.}
    \label{figure7}
\end{figure}

Additionally, the computed results of the QNFs varying with the charge $Q_0$ and the Lorentz-violating parameter $l$ are shown in Figs.~\ref{figure5}-\ref{figure7}. As we can see, for both the gravitationally‑dominated and electromagnetically‑dominated perturbations~\cite{Liu:2024oeq}, as $l$ increases, the absolute values of the imaginary parts of the QNFs of the gravitationally‑dominated and electromagnetically‑dominated fields decrease monotonically, while the real parts of the gravitationally‑dominated fields decrease monotonically but the real parts of the electromagnetically‑dominated fields increase monotonically.~As the electric charge $Q_0$ increases, both the real parts and the absolute imaginary parts of the QNFs increase.~As the angular momentum $L$ increases,~the real parts of the QNFs increase significantly,~while the imaginary parts remains nearly unchanged.
\begin{table}[h]
\centering
\begin{tabular}{|c|c|c|c|c|c|}
\hline
 $Q_0/2M$ & CFM &  AIM & WKB \\
\hline
&$2M\omega_R~~~-2M\omega_I$&$2M\omega_R~~~-2M\omega_I$&$2M\omega_R~~~~-2M\omega_I$ \\ \hline
 0   &1.19889~~0.18541&1.19889~~0.18541 &1.19889~~0.18541 \\
 0.1 &1.20207~~0.18558&1.20206~~0.18558& 1.20206~~0.18558\\
 0.2 &1.21413~~0.18613&1.21411~~0.18612& 1.21413~~0.18612 \\
 0.3 &1.24134~~0.18686& 1.24132~~0.18682 & 1.24132~~0.18683\\
 0.4 &1.29503~~0.18765&1.29510~~0.18763& 1.29506~~0.18624 \\
\hline
\end{tabular}
\caption{The fundamental frequencies of the QNMs of the gravitationally‑dominated perturbation for the charged black hole without Lorentz-violating effects, i.e.~$l=0$ (using CFM,~AIM, and WKB), for various values of the charge $Q_0$. The angular momentum $L=3$.}
\label{table4}
\end{table}
\begin{table}[h]
\centering
\begin{tabular}{|c|c|c|c|c|c|}
\hline
 $Q_0/2M$ & CFM &  AIM &WKB \\
\hline
&$2M\omega_R~~~-2M\omega_I$&$2M\omega_R~~~-2M\omega_I$&$2M\omega_R~~~~-2M\omega_I$ \\ \hline
 0   &1.31380~~0.19123&1.31380~~0.19123 &1.31380~~0.19123 \\
 0.1 &1.32874~~0.19194&1.32873~~0.19195 &1.32874~~0.19194\\
 0.2 &1.37457~~0.19394&1.37457~~0.19394&1.37456~~0.19395 \\
 0.3 &1.45838~~0.19673&1.45838~~0.19673 &1.45837~~0.19674\\
 0.4 &1.60573~~0.19823&1.58957~~0.18843 & 1.57881~~1.87740 \\
\hline
\end{tabular}
\caption{The fundamental frequencies of the QNMs of the electromagnetically‑dominated perturbation for the charged black hole without Lorentz-violating effects, i.e.~$l=0$ (using the CFM,~AIM and WKB), for various values of the charge $Q_0$. The angular momentum $L=3$.}
\label{table5}
\end{table}
\section{Conclusion}\label{V}

In the context of bumblebee gravity,~exact solutions to the field equations~\eqref{ds}-\eqref{Sr},~describing charged spherically symmetric black holes, have been presented~\cite{Liu:2024axg}.~Owing to the nonminimal coupling between the bumblebee field and the gravitational field, Lorentz symmetry is spontaneously broken in this framework.

In this paper, we studied the QNMs of the scalar perturbation and the gravito-electromagnetic coupled perturbation under the bumblebee black hole background. The perturbations can be expanded in terms of spherical harmonics due to the spherical symmetry of the background metric~\eqref{newds}. For the scalar perturbation, we directly obtained the QNFs by using the CFM and AIM.~For the gravito-electromagnetic perturbation, since the matrix equation~\eqref{master equation c} cannot be transformed into two separate equations,~we use the CFM to obtain the result. The relevant results were presented in Tabs.~\ref{table1}-\ref{table7} and Figs.~\ref{figure1}-\ref{figure7}.~It is observed that,~as the Lorentz-violating parameter $l$ increases, the absolute value of the imaginary parts of the QNFs for all three perturbation types decrease. The real parts of the QNFs for the electromagnetically‑dominated field slightly increases, while the real parts of the QNFs for both the scalar and gravitationally‑dominated perturbations decrease.~As the electric charge $Q_0$ increases, the absolute values of the imaginary parts of the QNFs for the scalar field remains nearly constant, both the imaginary parts of the QNFs for the other two perturbation types and the real parts of the QNFs for all three perturbations increase.~As the angular momentum $L$ increases,~the real parts of the QNFs increases significantly,~while the imaginary parts remains nearly unchanged.

We have only studied the quadratic potential with $\Lambda=0$ in this paper.~We hope to gain more understanding of black holes that include the cosmological constant and take into account the spin of black holes.~A relevant research will be carried out in the future.  

\acknowledgments\label{VI}

We hereby extend our sincere gratitude to Y.-M.~Ma for her assistance in running the numerical program,~which has significantly reduced the computation time,~and to Y.-F. He for sorting out the content of the references. ~This work is supported in part by the National Key Research and Development Program of China (Grant No. 2021YFC2203003), the National Natural Science Foundation of China (Grants No. 12475056, No. 12205129, and No. 12247101), Gansu Province’s Top Leading Talent Support Plan, the Natural Science Foundation of Gansu Province (No. 22JR5RA389), and the ‘111 Center’ under Grant No. B20063.
\begin{table}[h]
\centering
\begin{tabular}{|c|c|c|c|c|c|}
\hline
 $Q_0/2M$ & CFM &  AIM & WKB \\
\hline
&$2M\omega_R~~~-2M\omega_I$&$2M\omega_R~~~-2M\omega_I$&$2M\omega_R~~~~-2M\omega_I$ \\ \hline
 0   &1.61836~~0.18833& 1.49313~~1.81217 &1.61836~~0.18833 \\
 0.1 &1.62269~~0.18850& 1.44617~~2.28771 &1.62269~~0.18850\\
 0.2 &1.64041~~0.18907& 1.44497~~2.77423 &1.64037~~0.18922 \\
 0.3 &1.68111~~0.18987& 1.44427~~3.74354 &1.68102~~0.19078\\
 0.4 &1.76103~~0.19070& 1.49105~~5.99471 &1.75778~~0.20098 \\
\hline
\end{tabular}
\caption{The fundamental frequencies of the QNMs of the gravitationally‑dominated perturbation for the charged black hole without Lorentz-violating effects, i.e.~$l=0$ (using CFM,~AIM, and WKB), for various values of the charge $Q_0$. The angular momentum $L=4$.}
\label{table6}
\end{table}
\begin{table}[h]
\centering
\begin{tabular}{|c|c|c|c|c|c|}
\hline
 $Q_0/2M$ & CFM &  AIM &WKB \\
\hline
&$2M\omega_R~~~-2M\omega_I$&$2M\omega_R~~~-2M\omega_I$&$2M\omega_R~~~~-2M\omega_I$ \\ \hline
 0   &1.70619~~0.19172&1.42254~~2.27000 & 1.70619~~0.19172 \\
 0.1 &1.72520~~0.19242& 1.44490~~2.27618 &1.72520~~0.19242\\
 0.2 &1.78199~~0.19432& 1.42617~~2.76861 &1.78167~~0.19457 \\
 0.3 &1.88385~~0.19689& 0.94689~~3.46393 & 1.88266~~0.19709\\
 0.4 &2.06081~~0.19809& 1.49105~~5.99471 &2.05788~~0.19945\\
\hline
\end{tabular}
\caption{The fundamental frequencies of the QNMs of the electromagnetically‑dominated perturbation for the charged black hole without Lorentz-violating effects, i.e.~$l=0$ (using CFM,~AIM and WKB), for various values of the charge $Q_0$. The angular momentum $L=4$.}
\label{table7}
\end{table}
\begin{table}[h]
\centering
\begin{tabular}{|c|c|c|c|c|c|}
\hline
 $l$ & Gravitational & Electromagnetic \\
\hline
&~~$2M\omega_R~~~~~~~~~-2M\omega_I$~~~~~~~~~~$R$~&$2M\omega_R~~~~~~~~~-2M\omega_I$~~~~~~~~~~$R$~~\\ \hline
 0   &~~0.74959~~~~~~~~~~0.17819~~~~~~~~2.93393&0.92594~~~~~~~~~~0.19071~~~~~~~~0.03188 \\ \hline
 0.01 &~~0.74598~~~~~~~~~~0.17731~~~~~~~~2.57746&0.92615~~~~~~~~~~-0.18980~~~~~~~~0.03477
 \\ \hline 
 0.02 &~~0.74241~~~~~~~~~~0.17644~~~~~~~~2.21079&0.92635~~~~~~~~~~0.18889~~~~~~~~0.03767 \\ \hline
 0.03 &~~0.73890~~~~~~~~~~0.17559~~~~~~~~1.83260& 0.92657~~~~~~~~~~0.18799~~~~~~~~0.04072 \\ \hline
 0.04 &~~0.73543~~~~~~~~~~0.17474~~~~~~~~1.44689& 0.92676~~~~~~~~~~0.18711~~~~~~~~0.04426 \\ \hline
\end{tabular}
\caption{Testing the electromagnetic and gravitational branches of the QNMs varying with the Lorentz parameter $l$.}
\label{table8}
\end{table}

\appendix
\section{explicit expressions for CFM}\label{A}
In this appendix, we provide explicit expressions for the coefficients involved in Eq.~\eqref{iteration relation}.~To prevent the overall formula from becoming overly cumbersome, we default to taking $2M=1$,~and Appendix \ref{B} is handled in the same way.~Here,~the coefficients are given by

\begin{equation}
\begin{aligned}
\alpha_n &= (n+1)r_2^2\left[(n+1)\left(r_{2}-r_{1}\right)-2i(1+l)r_{2}^{2}\omega\right], \\
\beta_n &= r_2 \Bigl[ r_2^2 \bigl(6 l^2 r_2^2 \omega^2 + 2 l^2 r_2 \omega^2 - (1+l) L^2 - (1+l) L + 2 i (l+1) n (4 r_2+1) \omega \nonumber \\
&\quad + 12 l r_2^2 \omega^2 + 4 l r_2 \omega^2 + 3 i l r_2 \omega + i l \omega - 3 n^2 - 2 n + 6 r_2^2 \omega^2 + 2 r_2 \omega^2 + 3 i r_2 \omega + i \omega - 1 \bigr) \nonumber \\
&\quad + r_2 r_1 \bigl((l+1) L^2 + (l+1) L - 4 i n (l \omega + \omega + i) - 2 (l+1)^2 r_2 (r_2+1) \omega^2 \nonumber \\
&\quad - 2 i (1+l) (1+2r_2) \omega + n^2 + 2 \bigr)+ i r_1^2 \bigl(2 n [(1+l) (1+r_2) \omega + i] + (1+l) (1+r_2) \omega - 2 i n^2 + i \bigr) \Bigr],  \\
\gamma_n &= r_2^2 r_1 \bigl( -13 l^2 r_2^2 \omega^2 + 8 l^2 r_2 \omega^2 + 3 l^2 \omega^2 + (l+1) L^2 + (l+1) L- 6 i (l+1) n (2 r_2-1) \omega \nonumber \\
&\quad - 26 l r_2^2 \omega^2 + 16 l r_2 \omega^2 + 16 i l r_2 \omega + 6 l \omega^2 + 2 i l \omega + 3 n^2 - 12 n- 13 r_2^2 \omega^2 + 8 r_2 \omega^2 + 16 i r_2 \omega \nonumber \\
&\quad + 3 \omega^2 + 2 i \omega + 4 \bigr) -r_2^3 \bigl( 5 l^2 r_2^2 \omega^2 + 6 l^2 r_2 \omega^2 + l^2 \omega^2 - (l+1) L^2 - (l+1) L+ 4 i (l+1) n (2 r_2+1) \omega \nonumber \\
&\quad + 10 l r_2^2 \omega^2 + 12 l r_2 \omega^2 - 3 i l r_2 \omega + 2 l \omega^2 - i l \omega - 3 n^2 + 2 n+ 5 r_2^2 \omega^2 + 6 r_2 \omega^2 - 3 i r_2 \omega \nonumber \\
&\quad + \omega^2 - i \omega - 1 \bigr) + r_2 r_1^2 \bigl( l^2 r_2^2 \omega^2 - 4 l^2 r_2 \omega^2 - 3 l^2 \omega^2 - 2 (l+1) L^2 - 2 (l+1) L+ 4 i (l+1) (n-1) r_2 \omega \nonumber \\
&\quad + 2 l r_2^2 \omega^2 - 8 l r_2 \omega^2 + i l r_2 \omega - 6 l \omega^2 - 7 i l \omega - 5 (n-1)^2+ r_2^2 \omega^2 - 4 r_2 \omega^2 + i r_2 \omega - 3 \omega^2 - 7 i \omega + 4 \bigr) \nonumber \\
&\quad + r_1^3 \bigl( (n-1) \left[2 - 2 i (l+1) (2 r_2+1) \omega\right] + (l+1)^2 (r_2+1)^2 \omega^2+ 2 i (l+1) \omega - (n-1)^2 - 1 \bigr),  \\
\theta_n &= r_2 r_1^2 \bigl( 9 l^2 r_2^2 \omega^2 - 10 l^2 r_2 \omega^2 + 3 l^2 \omega^2 + (l+1) L^2 + (l+1) L- 6 i (n-2) (l \omega + \omega + i) \nonumber \\
&\quad + 18 l r_2^2 \omega^2 - 20 l r_2 \omega^2 - 5 i l r_2 \omega + 6 l \omega^2 + 8 i l \omega + 3 (n-2)^2+ 9 r_2^2 \omega^2 - 10 r_2 \omega^2 - 5 i r_2 \omega \nonumber \\
&\quad + 3 \omega^2 + 8 i \omega - 5 \bigr) +r_1^3 \bigl( 5 l^2 r_2^2 \omega^2 + 6 l^2 r_2 \omega^2 - l^2 \omega^2 + (l+1) L^2 + (l+1) L+ 4 i (n-2) (l \omega + \omega + i) \nonumber \\
&\quad + 10 l r_2^2 \omega^2 + 12 l r_2 \omega^2 + 2 i l r_2 \omega - 2 l \omega^2 - 3 i l \omega + 3 (n-2)^2+ 5 r_2^2 \omega^2 + 6 r_2 \omega^2 + 2 i r_2 \omega \nonumber \\
&\quad - \omega^2 - 3 i \omega + 2 \bigr) + r_2^2 r_1 \bigl( 11 l^2 r_2^2 \omega^2 + 4 l^2 r_2 \omega^2 - 3 l^2 \omega^2 - 2 (l+1) L^2 + 16 i (l+1) (n-2) r_2 \omega \nonumber \\
&\quad - 2 (l+1) L+ 22 l r_2^2 \omega^2 + 8 l r_2 \omega^2 + 2 i l r_2 \omega - 6 l \omega^2 - 7 i l \omega - 5 (n-2)^2+ 11 r_2^2 \omega^2 + 4 r_2 \omega^2 + 2 i r_2 \omega \nonumber \\
&\quad - 3 \omega^2 - 7 i \omega + 4 \bigr) - \left((l+1) r_1^4 \omega \left[ 2 (l+1) (r_2+1) \omega - 2 i (n-2) + i \right]\right) \nonumber \\
&\quad +r_2^3 \left[ (l+1) (r_2+1) \omega + i (n-2) + i \right]^2, \\
\mu_n &= r_1 \Biggl[ -r_1^2 \biggl( 3 l^2 r_2^2 \omega^2 + (l+1) L^2 + (l+1) L- 2 i (l+1) (n-3) (2 r_2-1) \omega + 6 l r_2^2 \omega^2 \nonumber \\
&\quad + 2 i l r_2 \omega - i l \omega + 3 (n-3)^2 - 2 (n-3)+ 3 r_2^2 \omega^2 + 2 i r_2 \omega - i \omega + 1 \biggr) \nonumber \\
&\quad + r_2 r_1 \biggl( -7 l^2 r_2^2 \omega^2 + 2 l^2 r_2 \omega^2 + (l+1) L^2 + (l+1) L- 4 i (l+1) (n-3) (2 r_2-1) \omega - 14 l r_2^2 \omega^2  \nonumber \\
&\quad + 4 l r_2 \omega^2+ 3 i l r_2 \omega - 2 i l \omega + (n-3)^2 - 4 (n-3)- 7 r_2^2 \omega^2 + 2 r_2 \omega^2 + 3 i r_2 \omega - 2 i \omega + 2 \biggr) \nonumber \\
&\quad -r_2^2 \biggl( 2 i (l+1) (n-3) (2 r_2+1) \omega + 2 (l+1)^2 r_2 (r_2+1) \omega^2+ i (l+1) (2 r_2-1) \omega - 2 (n-3)^2\biggl)\nonumber \\ &\quad+ (2 n-7)r_2^2   
 -(l+1) r_1^3 \omega \left[ 5 (l+1) r_2 \omega + 2 i (n-3) - i \right]+(l+1)^2 r_1^4 \omega^2 \Biggr], \\
\nu_n &= r_1^2 \biggl[2 i (l+1) (n-4) r_2^2 \omega+(l+1)^2 \left(r_1^3 + r_2 r_1^2 + r_2^2 r_1 + r_2^3\right) \omega^2+ (n-4)^2 \left(r_1 - r_2\right)\biggr]. 
\end{aligned}
\end{equation}
\section{explicit expressions for AIM}\label{B}

In this appendix, we provide explicit expressions for the coefficients involved in Eq.~\eqref{ODE1},~we rewrite this equation as
\begin{equation}
\psi''(\xi)=\lambda_0(\xi)\psi'(\xi)+s_0(\xi)\psi(\xi), 
\end{equation}
where,~the coefficients are given by
\begin{equation}
\begin{aligned}
\lambda_0(\xi) &= \frac{1}{(\xi -1)^2 \xi (r_2-r_1) ((\xi -1) r_1+r_2)}\Bigr(-(\xi -1) r_1^2 \bigl[2 \xi^2 ((l+1) \omega + 2i) \nonumber \\
&\quad -\xi (2(l+1)(r_2+1)\omega + 5i) + i\bigr]+2 r_1 r_2 \bigl[\xi^3 ((l+1)(r_2+1)\omega + 2i) \nonumber \\
&\quad -\xi^2 ((l+1)(4r_2+3)\omega + 6i)+\xi ((l+1)(5r_2+2)\omega + 5i)-(l+1)r_2\omega - i\bigr] \nonumber \\
&\quad +r_2^2 \bigl[\xi^2 (2(l+1)(r_2+1)\omega + 3i)-2\xi ((l+1)(3r_2+1)\omega + 2i)+2(l+1)r_2\omega + i\bigr]\Bigr), \\
s_0(\xi) &= -\frac{(l+1)^2 r_2^4 \omega ^2-(\xi -1)^2 \xi  \left((\xi -1) r_1+r_2\right) \left((l+1) L (L+1) r_2-(\xi -1) \left((2 \xi -1) r_1+r_2\right)\right)}{(\xi -1)^4 \xi ^2 \left((\xi -1) r_1+r_2\right)^2} \nonumber \\
&\quad +\frac{2 i (l+1) \omega }{(\xi -1)^2}+\frac{(l+1) \omega  (l \omega +\omega -i)}{(\xi -1)^2}+\frac{(l+1) r_2^2 \omega  \left(r_2 \left((l+1) r_2 \omega -i\right)+i r_1\right)}{\xi ^2 \left(r_1-r_2\right)^2} \nonumber \\
&\quad -\frac{i (l+1) r_2^2 \omega  \left((2 \xi -1) r_1+r_2\right)}{\xi ^2 \left(r_1-r_2\right) \left((\xi -1) r_1+r_2\right)}+\frac{2 (l+1)^2 r_2^3 \omega ^2}{(\xi -1)^2 \xi  \left(r_1-r_2\right)}+\frac{(l+1)^2 r_2^2 \omega ^2}{(\xi -1)^4} \nonumber \\
&\quad -\frac{2 (l+1)^2 r_2^2 \omega ^2}{(\xi -1) \xi  \left(r_1-r_2\right)}-\frac{(l+1)^2 r_2 \omega ^2}{(\xi -1)^3}-\frac{2 i (l+1) r_2^2 \omega }{(\xi -1) \xi  \left(r_1-r_2\right)}-\frac{2 i (l+1) r_2 \omega }{(\xi -1)^3} \nonumber \\
&\quad -\frac{(l+1) r_2 \omega  (l \omega +\omega -2 i)}{(\xi -1)^3}-\frac{i (l+1) r_2 \omega  \left((2 \xi -1) r_1+r_2\right)}{(\xi -1)^2 \xi  \left((\xi -1) r_1+r_2\right)}+\frac{i (l+1) \omega  \left((2 \xi -1) r_1+r_2\right)}{(\xi -1) \xi  \left((\xi -1) r_1+r_2\right)}.
\end{aligned}
\end{equation}

\bibliographystyle{JHEP}
\bibliography{ref}

\end{document}